\documentclass{article}

\PassOptionsToPackage{numbers, compress}{natbib}
\usepackage[final]{neurips_2021}

\usepackage[utf8]{inputenc} % allow utf-8 input
\usepackage[T1]{fontenc}    % use 8-bit T1 fonts
\usepackage{url}            % simple URL typesetting
\usepackage{booktabs}       % professional-quality tables
\usepackage{amsfonts}       % blackboard math symbols
\usepackage{nicefrac}       % compact symbols for 1/2, etc.
\usepackage{microtype}      % microtypography
\usepackage{xcolor}         % colors
\usepackage{graphicx} 
\usepackage{amsmath}
\DeclareMathOperator*{\argmax}{arg\,max}

\usepackage{kotex}
\usepackage{caption}
\usepackage{wrapfig}

\usepackage[colorlinks]{hyperref}
\definecolor{pinegreen}{rgb}{0.0, 0.47, 0.44}
\definecolor{cornellred}{rgb}{0.7, 0.11, 0.11}
\definecolor{cadmiumgreen}{rgb}{0.0, 0.42, 0.24}

\usepackage{xspace}
\setcitestyle{square}

\begin{document}
\title{Hit and Lead Discovery with Explorative RL and Fragment-based Molecule Generation}

\author{
  Soojung Yang \thanks{Currently at MIT.} \\
  AITRICS\\
  \texttt{soojungy@mit.edu} \\
   \And
   Doyeong Hwang \thanks{Currently at Onepredict.} \\
   AITRICS \\
   \texttt{desertbeagle11@gmail.com} \\
   \AND
  Seul Lee \\
  KAIST \\
  \texttt{ellenlee7890@gmail.com} \\
   \And
  Seongok Ryu \thanks{Currently at Galux inc.} \\
  AITRICS \\
   \texttt{seongokryu@galux.co.kr} \\
   \And
   Sung Ju Hwang \\
   AITRICS, KAIST  \\
   \texttt{sjhwang82@kaist.ac.kr} \\
}

\maketitle

\begin{abstract}
Recently, utilizing reinforcement learning (RL) to generate molecules with desired properties has been highlighted as a promising strategy for drug design. A molecular docking program -- a physical simulation that estimates protein-small molecule binding affinity -- can be an ideal reward scoring function for RL, as it is a straightforward proxy of the therapeutic potential. Still, two imminent challenges exist for this task. First, the models often fail to generate chemically realistic and pharmacochemically acceptable molecules. 
Second, the docking score optimization is a difficult exploration problem that involves many local optima and less smooth surfaces with respect to molecular structure. 
To tackle these challenges, we propose a novel RL framework that generates pharmacochemically acceptable molecules with large docking scores. 
Our method -- Fragment-based generative RL with Explorative Experience replay for Drug design (FREED) -- constrains the generated molecules to a realistic and qualified chemical space and effectively explores the space to find drugs by coupling our fragment-based generation method and a novel error-prioritized experience replay (PER). We also show that our model performs well on both \textit{de novo} and scaffold-based schemes. Our model produces molecules of higher quality compared to existing methods while achieving state-of-the-art performance on two of three targets in terms of the docking scores of the generated molecules. We further show with ablation studies that our method, predictive error-PER (FREED(PE)), significantly improves the model performance.
\end{abstract}

\section{Introduction}

Searching for ``hits", the molecules with desired therapeutic potentials, is a critical task in drug discovery. 
Instead of screening a library of countless potential candidates in a brute-force manner, designing drugs with sample-efficient generative models has been highlighted as a promising strategy. 
While many generative models for drug design are trained on the distribution of known active compounds \cite{jin2018junction, zhavoronkov2019deep, de2018molgan}, 
such models tend to produce molecules that are similar to that of the training dataset \cite{walters2020assessing}, which discourages finding novel molecules. 

\begin{figure}[h]
\hspace*{-0.1in}
  \centering
%   \fbox{\rule[-.5cm]{0cm}{4cm} \rule[-.5cm]{4cm}{0cm}}
  \includegraphics[width=1.0\linewidth]{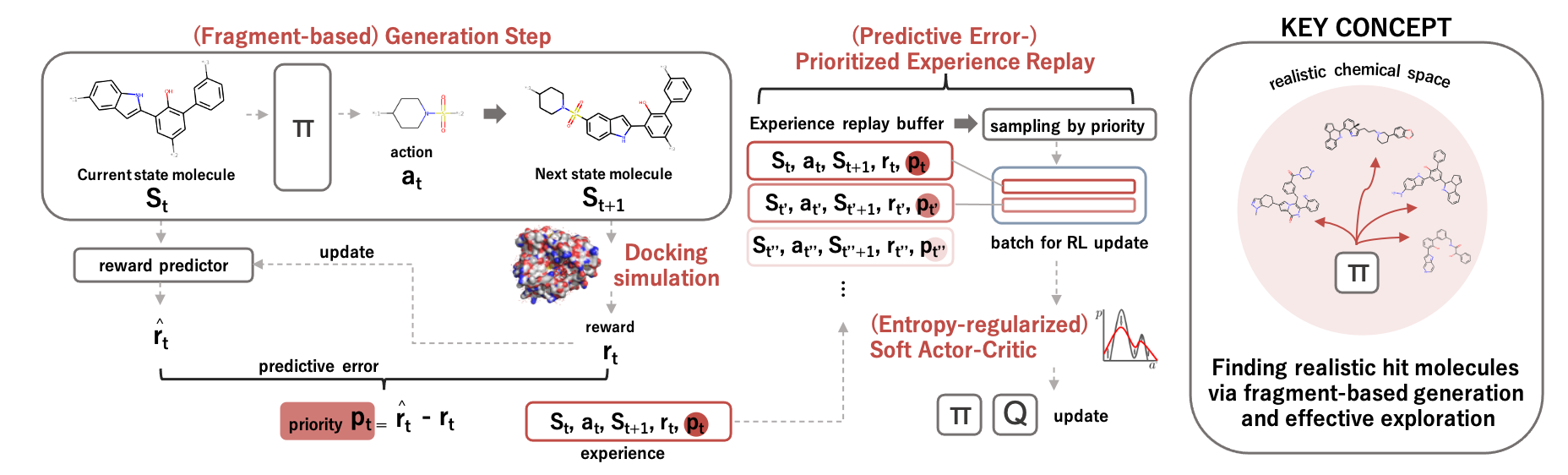}
  \caption{\textbf{Overview of our generative drug discovery method}. To find realistic `hit' molecules that have high docking scores, we combine fragment-based generation method with SAC and PER. This figure illustrates our version of PER where the priority of the experience is defined by predictive error.}
  \label{fig:concept1}
\end{figure}

In this light, reinforcement learning (RL) has been increasingly used for goal-directed molecular design, thanks to its exploration ability. 
Previous models have been assessed with relatively simple objectives, such as cLogP and QED, or estimated bioactivity scores predicted by auxiliary models \cite{zhou2019optimization, you2018graph, popova2018deep, blaschke2020memory}. 
However, high scores in those simple properties of molecules guarantee neither drug-likeness nor therapeutic potential, emphasizing the necessity of more relevant design objectives in generative tasks \cite{coley2021exploring, cieplinksi2020atleast}. 
The molecular docking method is a physical simulation that estimates the protein-small molecule binding affinity, a key measure of drug efficacy. As docking simulations are a more straightforward proxy of therapeutic potential, coupling RL with docking simulations would be a promising strategy. 
While the simplistic scores (e.g., cLogP, QED) are computed by a sum of local fragments' scores and are not a function of global molecular structure, making the optimization tasks relatively simple, docking score optimization is a more demanding exploration problem for RL agents. 
A change in docking score is nonlinear with the molecule's local structural changes, and a significant variance exists among the structures of high-scoring molecules, meaning that there exists many local optima \cite{coley2021exploring}.  

In addition, many previous RL models often suffer from generating unreal and inappropriate structures. 
Docking score maximization alone is not sufficient to qualify molecules as drug candidates, as the molecules should satisfy strong structural constraints such as chemical realisticness and pharmacochemical suitability. 
In other words, small molecule drug candidates should have enough steric stability to arrive at the target organ in the intended form (chemical realisticness), and they should not include any seriously reactive or toxic substructure (pharmacochemical suitability). 
The low quality of generated molecules can arise from a single improper addition of atoms and bonds, which would deteriorate the entire sanity of the structure. 
Since such `mistakes' are easy to occur, implicitly guiding the model (e.g., jointly training the model with QED optimization) cannot completely prevent the mistakes. 
Thus, explicitly restricting the generation space within realistic and qualified molecules by generating molecules as a combination of appropriate fragments can be a promising solution \cite{jin2018junction, jin2020hiervae, maziarz2021extend}.  

However, such a strong constraint in generation space would make the optimization problem of docking score even harder and render a higher probability of over-fitting to few solutions, urging a need for better exploration for RL agents. In this respect, we introduce a new framework, \textbf{Fragment-based generative RL with an Explorative Experience replay for Drug design (FREED)}, which encourages effective exploration while only allowing the generation of qualified molecules.

Our model generates molecules by attaching a chemically realistic and pharmacochemically acceptable fragment unit on a given state of molecules at each step\footnote{In this work, we widely define the term ``chemically realistic molecule" as a stable molecule and narrowly define it as a molecule that is an assembly of fragments that appear in the ZINC data. Also, we widely define the term ``inappropriate molecule/fragment" or ``pharmacochemically inacceptable molecule/fragment" as a molecule that has nonspecific toxicity or reactivity, and narrowly define it as a molecule that cannot pass through the three medicinal chemistry filters, which are Glaxo, PAINS, SureChEMBL filters.}. We enforce the model to form a new bond only on the attachment sites that are considered as appropriate at the fragment library preparation step. 
These strategies enable us to utilize medicinal chemistry prior knowledge and successfully constrain the molecule generation within the chemical space eligible for drug design.
We also explore several explorative algorithms based on curiosity-driven learning and prioritized experience replay (PER) \cite{schaul2015prioritized}.
We devise an effective novel PER method that defines priority as the novelty of the experiences estimated by the predictive error or uncertainty of the auxiliary reward predictor's outcome. With this method, we aim to avoid the lack of robustness of previous methods and encourage the exploration of diverse solutions. We provide an overall illustration of our framework in Figure \ref{fig:concept1}.  

Our main contributions can be summarized as follows:
\begin{itemize}
  \item We propose a novel RL framework that can be readily utilized to design qualified molecules of high therapeutic potential.  
  \item Our fragment-based generation method including connectivity-preserving fragmentation and augmentation allows our model to leverage Chemical prior knowledge.
  \item We propose novel explorative algorithms based on PER and show that they significantly improve the model performance. 
\end{itemize} 
\vspace{-0.1in}

\section{Related Works}
\label{related works}
\paragraph{SMILES-based and atom-based generation methods.}
SMILES-based methods \cite{olivecrona2017reinvent} are infeasible for scaffold-based tasks\footnote{A drug design scheme where the drug candidates are designed from the given scaffold.} since molecular structures can substantially change through the sequential extension of the SMILES strings. Also, as explained in the Introduction, atom-based generation methods such as You et al.'s GCPN \cite{you2018graph} inherently suffer from unrealistic generated molecules. Thus, we focus our discussion on motif-based generation methods.   
\vspace{-0.1in}

\paragraph{Motif-based molecular generation methods.}
A number of previous works \cite{jin2018junction, jin2020hiervae, maziarz2021extend} have investigated similar motif-based molecule generation based on the variational autoencoders (VAE). JT-VAE and HierVAE \cite{jin2018junction, jin2020hiervae} decompose and reconstruct molecules into a tree structure of motifs. These models might not be compatible with the scaffold-based generation, since a latent vector from their encoders depends on a motif-wise decomposition order which is irrelevant information for docking score that may bias the subsequent generation \cite{maziarz2021extend}. Maziarz et al. \cite{maziarz2021extend} also propose a VAE-based model which encodes a given scaffold with graph neural networks (GNNs) and decodes the motif-adding actions to produce an extended molecule. While Maziarz et al.'s motif-adding actions resemble our generation steps, we additionally introduce connectivity-preserving fragmentation and augmentation procedure which helps our model generate molecules of better quality.  

Coupling an RL policy network with fragment-based generation holds general advantages compared to VAE-based methods. For example, RL models do not need to be trained on reconstruction tasks which might restrict the diversity of generated molecules. Our work is one of the earliest applications of RL with a fragment-based molecular generation method. While Ståhl et al.'s DeepFMPO \cite{stahl2019deepfmpo} is also an application of RL with a fragment-based molecular generation method, DeepFMPO is designed to introduce only slight modifications to the given template molecules, which would make it inappropriate for the \textit{de novo} drug design. Moreover, while DeepFMPO's generation procedure cannot change the connectivity of the fragments of the given template, our method is free to explore various connectivity. 
\vspace{-0.1in}

\paragraph{Docking as a reward function of RL.}
Studies on docking score optimization task has started very recently. Jeon et al. \cite{jeon2020morld} developed MORLD, a atom-based generative model guided by MolDQN algorithm \cite{zhou2019moldqn}. Olivecrona et al. \cite{cieplinksi2020atleast} and Thomas et al. \cite{bender2021gpcr} utilized REINVENT\cite{olivecrona2017reinvent}, a simplified molecular-input line-entry system (SMILES)-based model generative model guided by improved REINFORCE algorithm \cite{williams1992reinforce} to generate hit molecules.
\vspace{-0.1in}

\paragraph{RL algorithms for hard-exploration problems.}
Our view is that, on a high level, there are two main approaches to achieve efficient exploration. The first one is to introduce the ``curiosity" or exploration bonus as intrinsic reward \cite{ryan2000intrinsic, oudeyer2009intrinsic, barto2013intrinsic} for loss optimization. Bellemare et al. \cite{bellemare2016unifying} first proposed a pseudo-count-based exploration bonus as an intrinsic reward. Pathak et al. \cite{pathak2017curiosity} defined curiosity as a distance between the true next state feature vector and the predicted estimate of the next state feature vector. Thiede et al. \cite{thiede2020curiosity} brought curiosity-driven learning into the molecular generation task. In Thiede et al., curiosity is defined as a distance between the true reward of the current state and the predicted estimate of the current state reward.
However, these ``curiosity-driven" intrinsic reward-based models sometimes fail to solve complex problems \cite{taiga2019benchmarking}. The failures are explained as a result of the RL agent's detachment or derailment from the optimal solutions \cite{ecoffet2019go}.

The other approach of solving hard exploration problems is a sample-efficient use of experiences \cite{paine2019making, schaul2015prioritized}. Prioritized experience replay (PER) method introduced in Schaul et al. \cite{schaul2015prioritized} samples experiences that can give more information to the RL agent and thus have more `surprisal' - defined by the temporal-difference (TD) error - in higher probability. In a similar sense, self-imitation learning (SIL) introduced in Oh et al. \cite{oh2018self} samples only the `good' experiences where the actual reward is larger than the agent's value estimate (i.e., estimate from the value function or Q function). However, prioritized sampling based on the agent's value estimate is susceptible to outliers and spikes, which may lead to destabilizing the agent itself \cite{lee2020predictive}.
Moreover, our explorative algorithm aims to preserve sufficient diversity and find many possible solutions instead of finding the most efficient route to a single solution. Thus, we modify the formulation of priority using the estimates for sample novelty.

\begin{figure}[]
\hspace*{-0.1in}
  \centering
%   \fbox{\rule[-.5cm]{0cm}{4cm} \rule[-.5cm]{4cm}{0cm}}
  \includegraphics[width=1.0\linewidth]{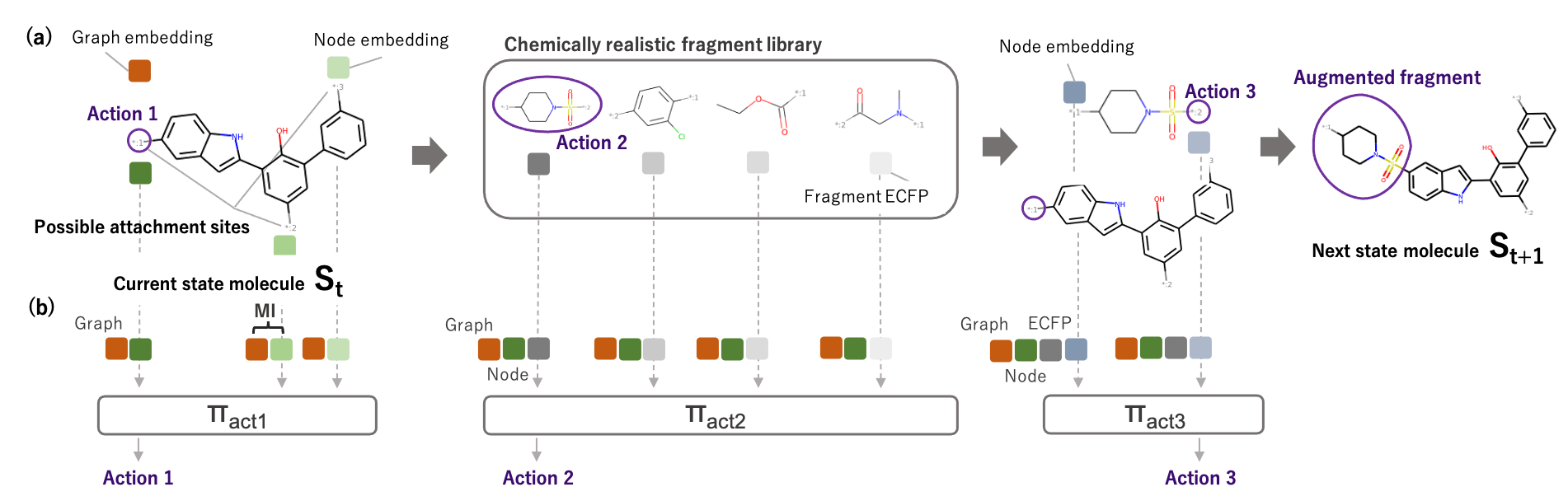}
  \caption{\textbf{Overview of our generation method \textbf{(a)} and policy network \textbf{(b)}}. Small colored squares represent a graph embedding of a molecule, node embedding of each attachment sitem and ECFP representation of each fragment. The graph or node embeddings and ECFP representations from the chosen actions are autoregressively passed onto the next action's policy network.}
  \label{fig:concept2}
\end{figure}

\vspace{-0.1in}
\section{Methods}
\label{methods}
\subsection{Generation method}
\paragraph{Generation steps.}
The key concept of our method is to generate high-quality molecules that bind well to a given target protein. 
In order to achieve this goal, we devise a fragment-based generation method in which molecules are designed as an assemble of chemically realistic and pharmacochemically acceptable fragments. 
Given a molecular state, our model chooses ``where to attach a new fragment (\textbf{Action 1})", ``what fragment to attach (\textbf{Action 2})", and ``where on a new fragment to form a new chemical bond (\textbf{Action 3})" in each step\footnote{Our model is designed to finish the episodes after four steps (in \textit{de novo} cases) or two steps (in scaffold-based cases). At the end of the episode, the model substitute all the remaining attachment sites with hydrogen atoms.}.
Note that the action ``where to form a new bond" makes our model compatible with docking score optimization since the scores depend on the three-dimensional arrangement of molecular substructures.  

\begin{figure}[]
  \centering
%   \fbox{\rule[-.5cm]{0cm}{4cm} \rule[-.5cm]{4cm}{0cm}}
  \includegraphics[width=0.8\linewidth]{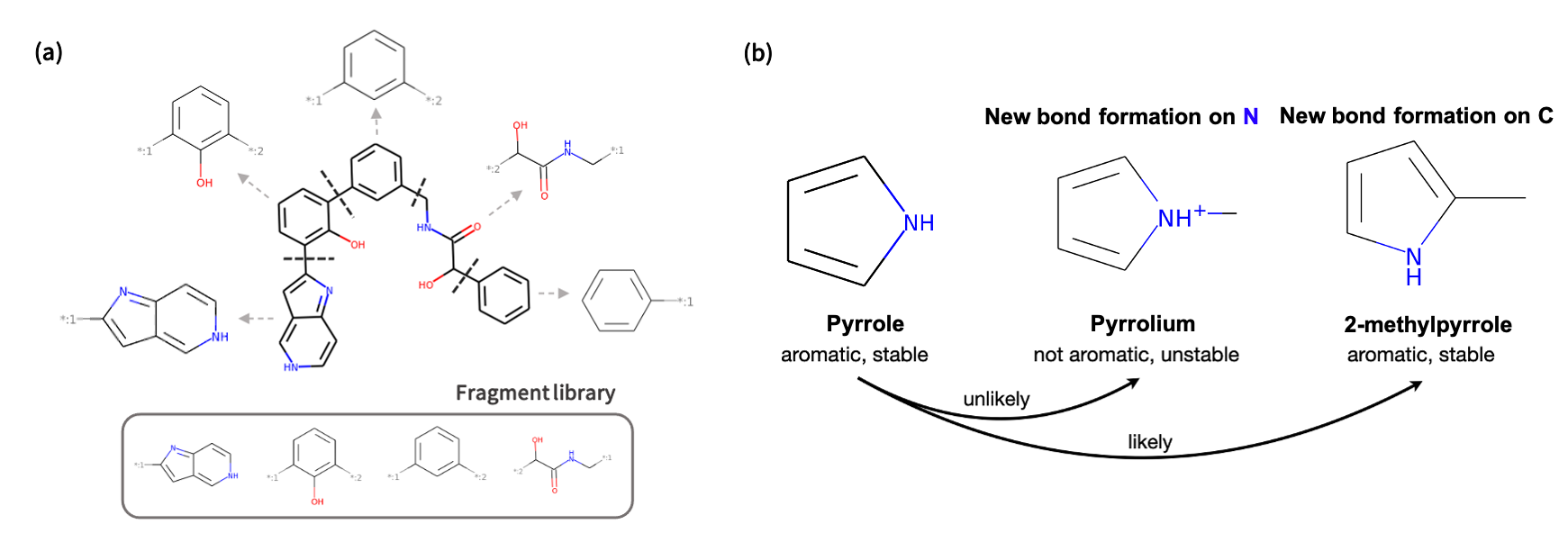}
  \caption{\textbf{Advantages of connectivity-preserving fragmentation.} \textbf{(a)} Substructure connectivity information is preserved as attachment sites in the fragmentation procedure. \textbf{(b)} Attaching a new bond to the nitrogen of a pyrrole ring would result in unstable molecule, breaking the aromaticity. Thus, pyrrole rings within an existing molecule are connected to other fragments by its carbon atoms, and preserving such an information would help us construct chemically realistic molecules.}
  \label{fig:concept3}
\end{figure}

\paragraph{Preserving fragment connectivity information in molecule generation.}
Another important feature of our method is harnessing the predefined connectivity information of the fragments and initial molecules. 
This feature allows our model to leverage chemists' expert knowledge in several ways.
The connectivity information arises in the fragmentation procedure, in which we follow CReM \cite{pavel2020crem} when the algorithm decomposes any arbitrary molecules into fragments while preserving the fragments' attachment sites as shown in Figure \ref{fig:concept3}\textbf{(a)}. 

In the GNN embedding phase, the attachment sites are considered as nodes like any other atoms in the molecule, while an atom type is exclusively assigned to the attachment sites. We also keep tracking the indices of the attachment sites as the states so that our policy can choose the next attachment site where a new fragment should be added (Action 1). Similarly, we keep the indices of the attachment sites of the fragments and use them throughout the training so that our policy can choose the attachment site from the fragment side (Action 3).

Start and end points of new bonds are restricted to the attachment sites of given fragments and molecules.
This restriction contributes to the chemical realisticness since the stability of the molecule depends on where the fragment is attached as illustrated in Figure \ref{fig:concept3}\textbf{(b)}. 
Also, we can utilize our prior knowledge of protein-ligand interaction by rationally assigning the attachment sites of the initial molecule (scaffold), which have been widely harnessed by medicinal chemists in a usual scaffold-based scenario.

\paragraph{Policy network.}
Our generation method is coupled with the policy networks which guide the model to generate hit molecules. We provide an overall illustration of our policy network in Figure \ref{fig:concept2}.  
Our state embedding network and policy network are designed as Markov models so that generation steps can take any arbitrary molecule as a current state and predict the next state, making the model more plausible for a scaffold-based generation. 

In our framework, each state vector $s_t$ represents a molecular structure at the $t$-th step, and each step is a sequence of three actions that define an augmentation of a fragment. Following the method from Hwang et al. \cite{hwang2020comprehensive}, the current state molecule is represented as an undirected graph $G$ which consists of a node feature matrix, an adjacency matrix, and an attachment site masking vector. 
A graph convolutional network (GCN) \cite{kipf2016semi}-based network provides node embeddings, $H$, which are then sum-aggregated to produce a graph embedding, $h_g$. Note that the same GCN encodes the current state molecular graph (for Action 1) and the fragments (for Action 3). 

Our policy network is an autoregressive process where Action 3 depends on Action 1 \& 2 and Action 2 depends on Action 1.
\begin{align}
\label{eqn:policy_network}
p^{\text{act1}} = & \pi_{\text{act1}}(Z^{\text{1st}}),  Z^{\text{1st}} = \text{MI}(h_g, H_\text{att})  \\
p^{\text{act2}} = & \pi_{\text{act2}}(Z^{\text{2nd}}),  Z^{\text{2nd}} = \text{MI}(z^{\text{1st}}_{\text{act1}}, \text{ECFP}(h_{g_{\text{cand}}})) \\
p^{\text{act3}} = & \pi_{\text{act3}}(Z^{\text{3rd}}),  Z^{\text{3rd}} = \text{MI}(z^{\text{2nd}}_{\text{act2}}, U_{\text{cand}})
\end{align}
where $H_\text{att}$ refers to the node embeddings of attachment sites.
For the first step of our policy network, $\pi_{\text{act1}}$ takes the multiplicative interactions (MI) \cite{jayakumar2019multiplicative} of the node embedding of each attachment site and the rows of the graph embedding of the molecule, $ z^{1st}_{i} \in Z^{1st}$, as inputs and predicts which attachment site should be chosen as Action 1. 
Since graph embeddings and node embeddings are defined in a heterogeneous space, we apply MI to fuse those two sources of information. 

Given Action 1, $\pi_{\text{act2}}$ takes the MI of $z^{\text{1st}}_{\text{act1}}$, the row of $Z^{\text{1st}}$ under index $act1$, and the RDKit \cite{rdkit} ECFP (Extended Connectivity Fingerprint) representation of each candidate fragment, $h_{g_{\text{cand}}}$ as inputs, 
and predicts which fragment should be chosen as Action 2. 
By taking $Z^{\text{1st}}$ as one of the inputs, $\pi_{\text{act2}}$ reflects on the context of the current state and Action 1. 

Finally, given Action 1 and 2, $\pi_{\text{act3}}$ takes the MI of $z^{\text{2nd}}_{\text{act2}}$
and the node embeddings of the chosen fragment's attachment sites, $U_{\text{cand}}$, as inputs and predicts which fragment attachment site should be chosen as Action 3.
Each of the three policy networks $\pi_{\text{act1}}, \pi_{\text{act2}}$ and $\pi_{\text{act3}}$ consists of three fully-connected layers with ReLU activations followed by a softmax layer to predict the probability of each action. In order to make gradient flow possible while sampling discrete actions from the probabilities, we implement the Gumbel-softmax reparameterization trick \cite{jang2016categorical}. 

\subsection{Explorative RL for the discovery of novel molecules}
\paragraph{Soft actor-critic.}
We employ \emph{soft actor-critic} (SAC), an off-policy actor-critic RL algorithm based on maximum entropy reinforcement learning \cite{haarnoja2018softog, haarnoja2018soft}, which is known to explore the space of state-action pairs more effectively than predecessors.
Assuming that our docking score optimization task requires more effective exploration than simplistic tasks, we chose SAC as our RL baseline algorithm.

\vspace{-0.05in}
\begin{align}
&\pi^{*} = \argmax_{\pi} \sum_{t} \mathbb{E}_{(s_t, a_t)\sim\rho_{\pi}} [r(s_t, a_t) + \alpha\mathcal{H}(\pi(\cdot|s_t))] \label{eqn:maximum_entropy}\\
\begin{split}
& \mathbb{E}_{s_{t}\sim \mathcal{D}}[\alpha \log \pi(a|s)] = \sum [\alpha \pi(a_{t}^{\text{act1}}, a_{t}^{\text{act2}}, a_{t}^{\text{act3}}|s_{t})\\
& \times(\log \pi(a_{t}^{\text{act1}}|s_{t}) + \log \pi(a_{t}^{\text{act2}}|s_{t}, a_{t}^{\text{act1}}) + \log \pi(a_{t}^{\text{act3}}|s_{t}, a_{t}^{\text{act1}}, a_{t}^{\text{act2}}))]
\label{eqn:entropy_decomposition}
\end{split}
\end{align}

SAC aims to attain optimal policy that satisfies \eqref{eqn:maximum_entropy}
where $\alpha$ is the temperature parameter balancing between exploration and exploitation of the agent, $\mathcal{H}(\pi(\cdot|s_t))$ is entropy of action probabilities given $s_t$, and $\rho_{\pi}$ is state-action transition distributions created from $\pi$. As we define $\text{act1}$, $\text{act2}$ and $\text{act3}$ autoregressively, the entropy regularization term defined in \cite{christodoulou2019soft} $\mathbb{E}_{s_{t}\sim \mathcal{D}}[\alpha \log \pi(a|s)]$ is decomposed into \eqref{eqn:entropy_decomposition}.

\paragraph{Explorative algorithms.}
To encourage exploration, we prioritize novel experiences during sampling batches for RL updates. We regard an experience as a novel experience if the agent has not visited the state before. Defining priority estimate function in the state space and not in the state-action space has been introduced for the molecular generative task in Thiede et al. \cite{thiede2020curiosity}. For novel states, the reward estimator would yield a high predictive error or high variance (Bayesian uncertainty).  In this regard, we train an auxiliary reward predictor consisting of a graph encoder and fully-connected layers that estimate a given state's reward (docking score). Then, we use the predictor's predictive error (L2 distance) or Bayesian uncertainty as a priority estimate. We name the former method \textbf{PER(PE)} and the latter method \textbf{PER(BU)}. The use of predictive error as a novelty estimate has been introduced for curiosity-driven learning \cite{thiede2020curiosity}, but our work is the first to apply this to PER.   

For PER(BU), we follow Kendall et al. \cite{kendall2017uncertainties} to obtain the Bayesian uncertainty of the prediction. We train the reward predictor network to estimate the reward's mean and variance (aleatoric uncertainty). Every layer in the network is MC dropout \cite{gal2016dropout} layer so that the predictor can provide the epistemic uncertainty. We add aleatoric and epistemic uncertainty to obtain the total uncertainty of the estimate.  

The reward predictor is optimized for every docking step, and we only optimize it based on final state transitions since docking scores are only computed for the final states. The reward predictor predicts the reward of any state, both intermediate and final. For PER(PE), when we compute the priority of a transition including an intermediate state, we use Q value as a substitute for the true docking score. After updating the policy network and Q function with loss multiplied by importance sampling (IS) weight, we recalculate and update the priority values of the transitions in the replay buffer. 

The following experiments section shows that PER with our priority estimate functions performs better than the previous methods.  

\section{Results and Analysis}
\subsection{Quantitative metrics}
In this section, we introduce quantitative metric scores we used to assess our model. For every metric, we repeated every experiment five times with five different random seeds and reported the mean and the standard deviation of the scores. Also, we calculated the scores when 3,000 molecules were generated and used to update the model during training.  
\paragraph{Quality score.} 
We report three widely used pharmacochemical filter scores -- Glaxo \cite{lane2006defining}, SureChEMBL \cite{papadatos2016surechembl}, PAINS \cite{baell2010new} -- as quality scores. The quality scores are defined as a ratio of accepted, valid molecules to total generated molecules, as the filters reject the compounds that contain functional groups inappropriate for drugs (i.e., toxic, reactive groups). The higher the quality scores, the higher the probability that the molecule will be an acceptable drug. We also report the ratio of valid molecules to total generated molecules (validity) and the ratio of unique molecules among valid generated molecules (uniqueness).  
\paragraph{Hit ratio.} We define hit ratio as a ratio of unique hit molecules to total generated molecules. We report the hit ratio to compare the model's ability to produce as many unique hit molecules in a given length of iterations, where we define \emph{hits} as molecules whose docking scores are greater than the median of known active molecules' docking scores. 
\paragraph{Top 5\% score.} We report the average score of the top 5\%-scored generated molecules to compare the model's ability to produce molecules with better docking scores.

\subsection{Quantitative performance benchmark}
In this section, we compare the quality of the generated molecules and the model performance with three baseline models, \textbf{MORLD} \cite{jeon2020morld}, \textbf{REINVENT} \cite{olivecrona2017reinvent}, and \textbf{HierVAE} \cite{jin2020hiervae}. Our model, \textbf{FREED(PE)}, is our fragment-based generation method coupled with SAC and PER(PE). MORLD and REINVENT are the models utilized for docking score optimization tasks in previous works \cite{jeon2020morld, cieplinksi2020atleast, bender2021gpcr}. HierVAE is a strong non-RL fragment-based molecular generative model. Since HierVAE is a distributional learning method, we train HierVAE in two schemes - `one-time' training on the known active molecules (\textbf{HierVAE}) and `active learning (AL)' training where we train the model once on the known actives and twice on the top-scoring molecules from the generated molecules (\textbf{HierVAE(AL)}). In both schemes, we initialized the models with the model pre-trained on ChEMBL.    
We trained the models for three carefully chosen protein targets fa7, parp1, and 5ht1b. The choice of protein targets and the specifics of experimental settings are described in Appendix \ref{sec:A1}.
For the experiments in this section, we use the small fragment library that includes 66 pharmacochemically acceptable fragments.

% \vspace{-0.05in}
\begin{table}[h]
\caption{\textbf{Quality scores of the models.} We trained our model and three baseline models with the target fa7 and computed quality scores of the first 3,000 molecules generated during training for each model. The two baseline models REINVENT and MORLD that are jointly trained to maximize filter scores are noted as `REINVENT w/ filter' and `MORLD w/ filter'. Standard deviation is given in brackets.}
\label{tab:quality}
% \vspace{-0.1in}
\centering
\begin{tabular}{llllll}
\toprule
          & Glaxo & SureChEMBL & PAINS & validity  & uniqueness \\
          \midrule
MORLD     & 0.561 (.009) & 0.131 (.013) & 0.805 (.013) & \textbf{1.000} (.000)  & \textbf{1.000} (.000) \\ 
MORLD w/ filter & 0.578 (.010) & 0.145 (.018) & 0.816 (.008) & \textbf{1.000} (.000)  & \textbf{1.000} (.001) \\
REINVENT & 0.773 (.023) & 0.667 (.030) & 0.769 (.022) & 0.813 (.024) & 0.988 (.008)    \\
REINVENT w/ filter & 0.832 (.034) & 0.747 (.040) & 0.842 (.034) & 0.872 (.028) & 0.990 (.007)  \\
HierVAE & 0.899 (.027) & 0.748 (.024) & 0.975 (.006) &  1.000 (.000) & 0.138 (.006)\\
HierVAE(AL) & 0.975 (.004) & 0.795 (.007) & 0.893 (.011) & 1.000 (.000) & 0.131 (.003) \\
\midrule
Ours: FREED(PE) & \textbf{0.996} (.001) & \textbf{0.808} (.049) & \textbf{0.991} (.002) & \textbf{1.000} (.000) & 0.723 (.135) \\
\bottomrule
\end{tabular}
\end{table}

\begin{figure}
\vspace{-0.2in}
  \centering
%   \fbox{\rule[-.5cm]{0cm}{4cm} \rule[-.5cm]{4cm}{0cm}}
  \includegraphics[width=0.9\linewidth]{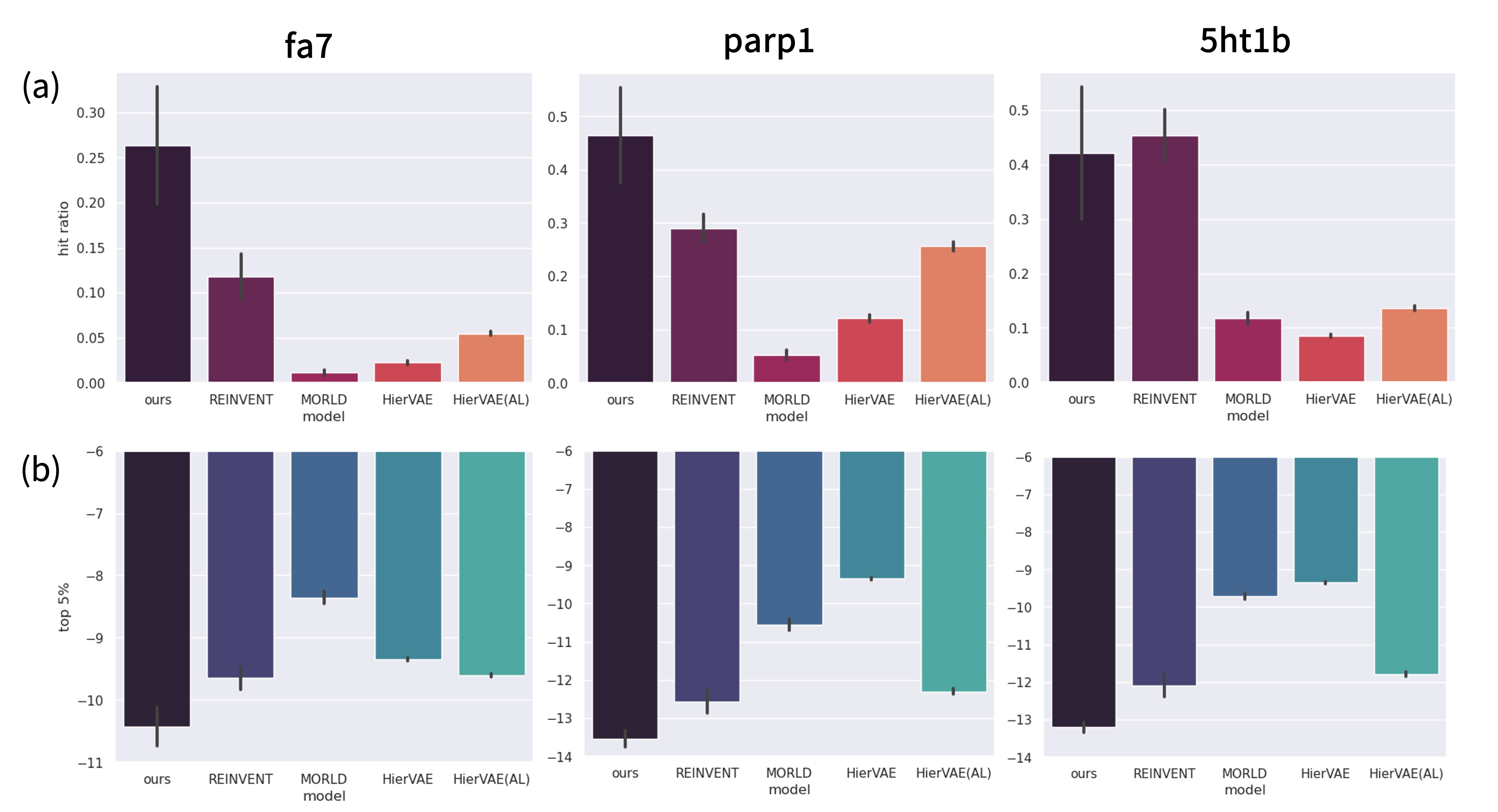}
  \caption{\textbf{Hit ratio and top 5\% score of our model FREED(PE), REINVENT, MORLD, HierVAE, and HierVAE(AL).} Standard deviation is given as error bars. Higher hit ratios and greater negative value of the top 5\% scores indicate better performance.}
  \label{fig:bar1}
\end{figure}
\vspace{-0.1in}

\paragraph{Quality scores of generated molecules.}
In real-world drug design, the generated molecules should be acceptable by pharmacochemical criteria. We excluded the fragments that are considered inappropriate by pharmacochemical filters to guarantee the quality of the generated molecules. Such an explicit constraint cannot be applied to atom-based (MORLD) or SMILES-based (REINVENT) methods.  

As shown in Table \ref{tab:quality}, our model mostly generated acceptable molecules while the other models showed the poor rate of acceptable molecules, confirming our approach's advantage in drug design. We also investigated whether one could improve the quality of the molecules generated from the baselines by using a multi-objective reward function. MORLD and REINVENT optimized in multi-objective with both docking scores and the quality scores, denoted as `MORLD w/ filter' and `REINVENT w/ filter', show improved quality scores compared to single-objective MORLD and REINVENT. However, the multi-objective reward method was not as effective as our fragment-based approach, strengthening our claim that the explicit constraints are the effective strategy. Such a trend was consistent for all three targets. 

HierVAE showed high quality scores, as the HierVAE fragment library itself had very few problematic substructures (See Appendix \ref{sec:A5} for details). Such a result substantiates the advantage of the explicit fragment-based approach. However, HierVAE showed low uniqueness, possibly due to the small size of training data for fine-tuning (\textasciitilde 1,200 known active molecules) and active learning (\textasciitilde 1,500 generated high-scoring molecules).   

We provide quality scores of our model (FREED(PE)) trained with the small library and the large library for all three protein targets in Table \ref{tab:quality_2_ap} and Table \ref{tab:quality_3_ap} of Appendix \ref{sec:A1}, respectively. A significant increase in uniqueness was observed when we used the large library, which implies that the low uniqueness of our model is due to the small size of the fragment library. Thus, we believe constructing a fragment library that is large enough to guarantee high uniqueness while only including pharmacochemically acceptable fragments will be the best strategy in production.

\vspace{-0.1in}
\paragraph{Docking score optimization result.}
Figure \ref{fig:bar1} shows the hit ratios and top 5\% scores of the generative models. Our model outperforms the other generative models MORLD, REINVENT, and HierVAE in terms of both hit ratio and top 5\% score, except for the hit ratio of the 5ht1b case. 
Such results show that our model's performance is superior or at least competitive to existing baselines, while our model exhibits many more practical advantages such as generating acceptable molecules, integrating chemist's expert knowledge by connectivity-preserving fragmentation and augmentation, and the feasibility in both \textit{de novo} and scaffold-based generation. 

% \vspace{0.2in}
\subsection{Ablation studies: explorative algorithms}
We also perform ablation studies of our algorithm to investigate the effect of our explorative algorithms. We used the larger library of 91 unfiltered fragments, as this section assesses the effect of the algorithms on the model's performance regardless of the pharmacochemical acceptability.

\begin{figure}[h]
\hspace*{-0.1in}
  \centering
%   \fbox{\rule[-.5cm]{0cm}{4cm} \rule[-.5cm]{4cm}{0cm}}
  \includegraphics[width=1.0\linewidth]{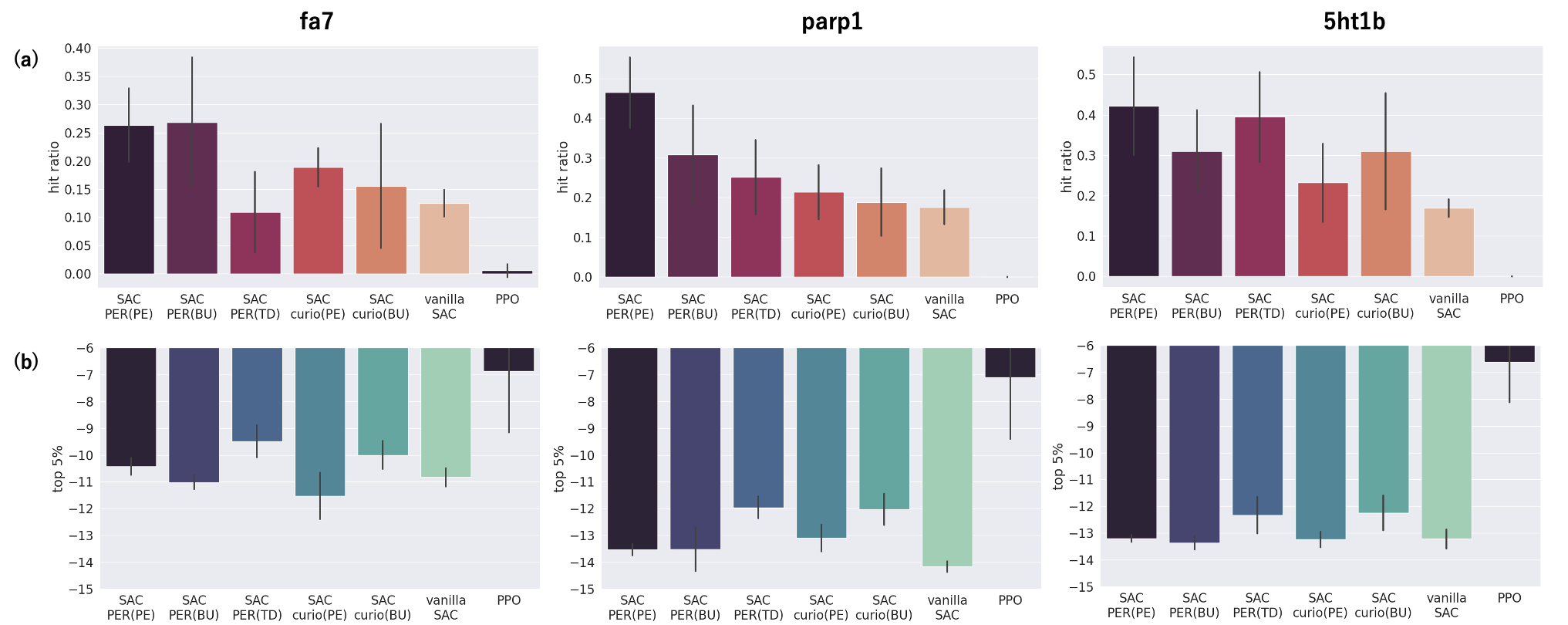}
  \caption{\textbf{Hit ratio and top 5\% score of ablation studies.} Models can be categorized by whether they use \{PER, curiosity-driven exploration(curio)\}, and whether they use \{predictive error from predictor(PE), Bayesian uncertainty(BU), and TD error from agent(TD)\} as priority or intrinsic reward. Standard deviation is given as error bars.}
  \label{fig:bar2}
\end{figure}

In Figure \ref{fig:bar2}, we observe that all SAC models with explorative algorithms performed better than the vanilla SAC model, while the PPO model showed the worst performance. FREED with both PE and BU outperformed curiosity-driven models with PE and BU, showing the effectiveness of our methods in our task compared to curiosity-driven explorations.
Moreover, our predictive error-PER method outperformed the TD error-PER method. We conjecture that such a result is due to 1) novelty-based experience prioritization encourages better exploration 2) leveraging an auxiliary priority predictor network makes PER training more robust than internal value estimate functions (Q function). We provide the significance analysis (one-tail paired t-test) of the result in Table \ref{tab:significance_ap} of Appendix \ref{sec:A1}.  

\begin{figure}[h]
\hspace*{-1cm}
  \centering
%   \fbox{\rule[-.5cm]{0cm}{4cm} \rule[-.5cm]{4cm}{0cm}}
  \includegraphics[width=1.0\linewidth]{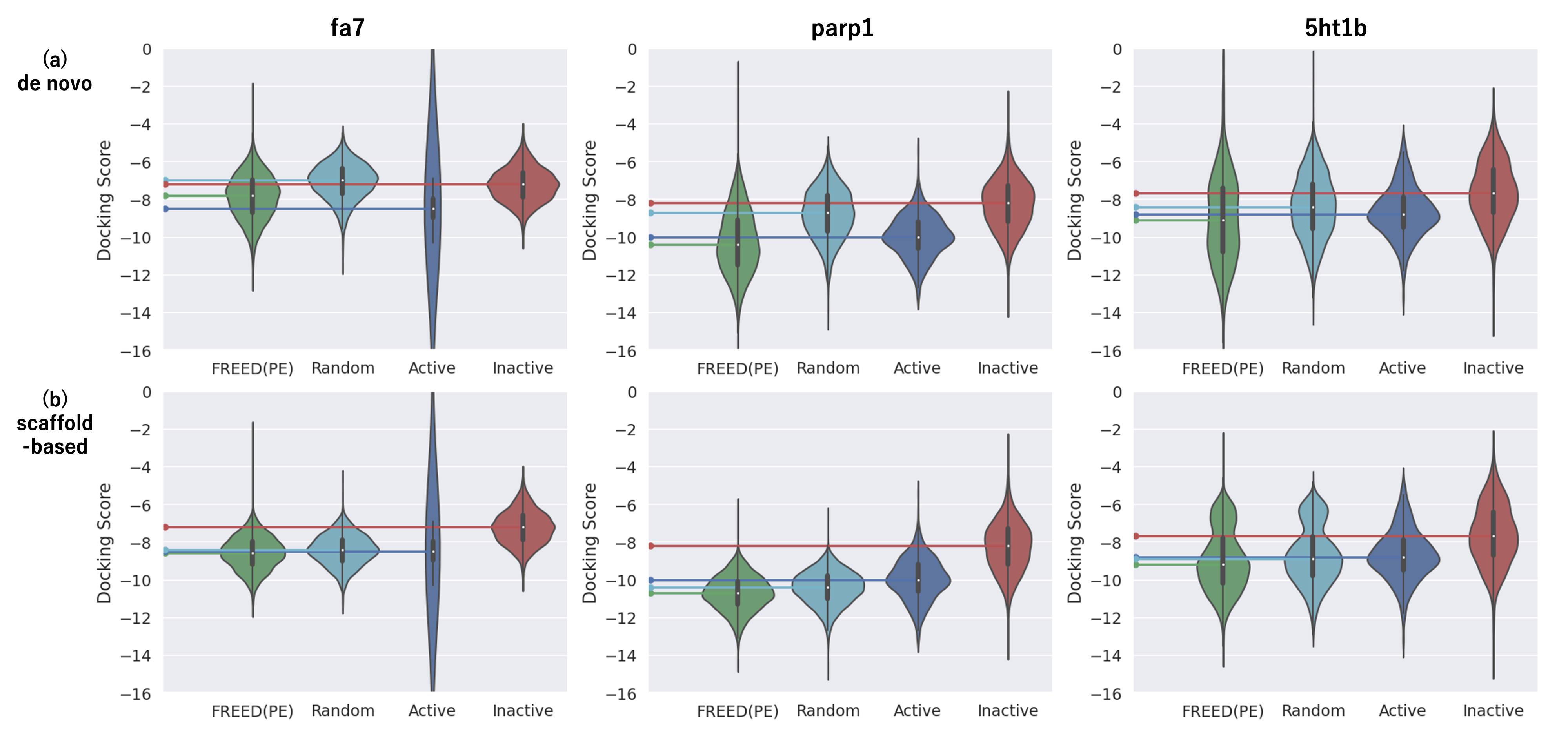}
  \caption{\textbf{Docking score distribution of the generated molecules.} Duplicate molecules were removed after gathering 3,000 molecules each from five random seed experiments. ``Random" molecules are generated by our fragment-based generation method without training the policy network. ``FREED(PE)" molecules are generated by the fragment-based generation method while training the policy network. We also plot known ``Active" and ``Inactive" molecules from DUD-E (fa7, parp1) or ChEMBL (5ht1b) datasets for comparison. Colored horizontal lines indicate the median of the corresponding distribution. \textbf{(a)} \textit{de novo} scenario \textbf{(b)} scaffold-based scenario}
  \label{fig:violin1}
\end{figure}
\vspace{-0.1in}

% \vspace{-0.1in}
\subsection{Case study on drug design}
In this section, we show the practicality of our framework on \textit{de novo} and scaffold-based drug design. We test FREED(PE) with our large fragment library which includes 91 fragments.  

\begin{figure}[h]
  \centering
%   \fbox{\rule[-.5cm]{0cm}{4cm} \rule[-.5cm]{4cm}{0cm}}
  \includegraphics[width=1.0\linewidth]{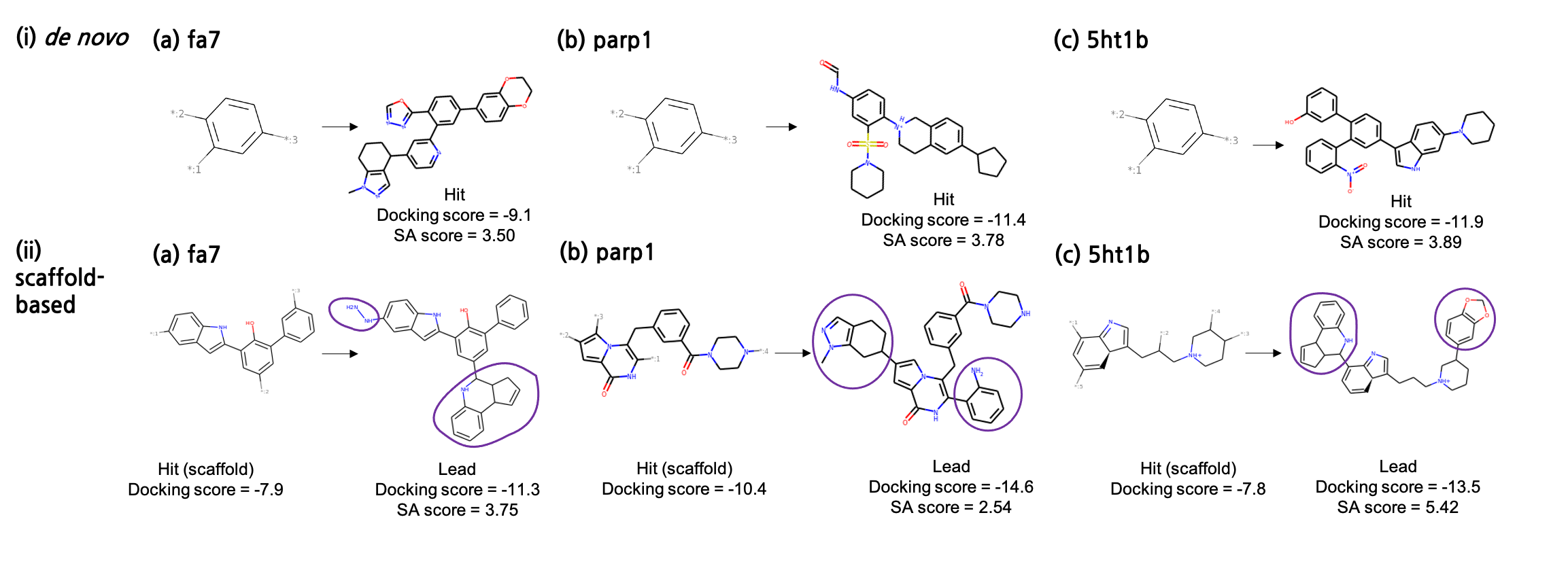}
  \caption{\textbf{Generated samples and their docking scores with our method, for \textit{de novo} \textbf{(i)} and scaffold-based scenario \textbf{(ii)}.} For each target, one of the high-scoring generated molecules is displayed with the initial molecule (benzene ring or scaffold). The purple line highlights the fragments augmented by the model in a scaffold-based generation. Numbers below the compounds are the docking scores and SA scores.}
  \label{fig:genmol}
\end{figure}

\paragraph{\textit{De novo} scenario.}
Figure \ref{fig:violin1} \textbf{(a)} shows the distribution of the generated molecules before (``random") and after (``FREED(PE)") optimizing the policy network. Our model was able to effectively generate molecules that have higher docking scores compared to the known active molecules. Figure \ref{fig:genmol} \textbf{(i)} shows the structure of each target's optimized molecules. 

\paragraph{Scaffold-based scenario.}
We validate our model on a scaffold-based scenario, where we attempt to improve docking scores by adding fragments to an initial scaffold molecule. Figure \ref{fig:violin1} \textbf{(b)} shows the distribution of the optimized molecules before (``random") and after (``FREED(PE)") training the policy network, with a scaffold of each target as an initial molecule. 

\vspace{-0.1in}

Figure \ref{fig:violin1}\textbf{(b)} highlights our model's ability to optimize a given scaffold to have a higher binding affinity with the target. Surprisingly, in Figure \ref{fig:violin1}\textbf{(b)}, even the molecules randomly optimized with our fragment-based generation algorithm show high docking scores when given the proper scaffold. This result implies the importance of scaffold in hit discovery and highlights our generative method's ability to to span the chemical space around the scaffold effectively. 

Figure \ref{fig:genmol}\textbf{(ii)} shows the structure of each target's scaffold and corresponding optimized molecules. We can see that the scaffold structures are well preserved in generated lead molecules. We provide an analysis of 3D docking poses of the scaffolds and generated lead molecules in Figure \ref{fig:all_pose_ap} and Figure \ref{fig:fa7_pose_ap} of Appendix \ref{sec:A1}.  

It is notable that our framework does not have to be specific for \textit{de novo} or scaffold-based scenarios, except for the initial molecule and number of fragments to be added. Since our model is fully Markovian, whether the initial molecule is a benzene ring or a scaffold does not affect the model's training.

\paragraph{Chemical realisticness of generated molecules.} In Figure \ref{fig:genmol}, we report the SA (synthetic accessibility) score of the molecules, which is a widely used metric that estimates ease of synthesis by penalizing the presence of non-standard structural features. The SA score distribution of the catalogue molecules of commercial compound providers has its mode around 3.0 \cite{ertl2019sa}. Accordingly, we can assume our generated molecules as reasonably synthesizable and thus chemically realistic.  

\vspace{-0.1in}
\section{Conclusion}
\vspace{-0.1in}
In this work, we developed FREED, a novel RL framework for real-world drug design that couples a fragment-based molecular generation strategy with a highly explorative RL algorithm to generate qualified hit molecules. Our model generates pharmacochemically acceptable molecules with high docking scores, significantly outperforming previous docking RL approaches. Our code is released at \url{https://github.com/AITRICS/FREED}. 

\textbf{Limitations and future work.} While our method does not explicitly account for the synthesizability of generated molecules, we believe forward synthesis-based methods \cite{gottipati2020navigate} can be complementary to ours. 
It would be able to combine our method and forward synthesis-based method by substituting our attachment site bond formation actions with chemical reactions. In this way, we can explicitly take synthesizability into account while providing the appropriate model inductive bias for docking score optimization. We leave such an improvement as future work.  

\textbf{Negative societal impacts.} If used maliciously, our framework can be utilized to generate harmful compounds such as biochemical weapons. Thus, conscious use of the AI model is required. 

\begin{ack}
We thank anonymous reviewers, Seung Hwan Hong, and Sihyun Yu for providing helpful feedback and suggestions. This work was supported by the National Research Foundation of Korea (NRF) grant funded by the project NRF2019M3E5D4065965.
\end{ack}

\bibliography{refer}

\newpage
\appendix
\section{Appendix}

\subsection{Additional experimental results}
\label{sec:A1}
We further introduce our additional experiments in this section. 

\paragraph{Training baseline models jointly with pharmacochemical filter scores.} 
In our main article, we compared our model FREED with baseline models REINVENT and MORLD. For fairer comparison of quality scores, we also performed multi-objective optimization of REINVENT and MORLD on both quality score (pharmacochemical filter score) and docking score as follows. 
\vspace{-0.05 in}
\begin{align}
\label{eqn:reward}
\text{total reward} = \text{docking score reward} + \text{pharmacochemical filter reward} * 0.5
\end{align}
Pharmacochemical filter scores, calculated as 1 if accepted and 0 if rejected by each filter, were multiplied by 0.5 and then added to docking score reward ($\text{docking score reward} = - \text{docking score}$) to obtain the total reward.  

Table \ref{tab:quality} of the main text shows that such an implicit method is not enough to achieve nearly perfect filter scores as our model did. Also, as shown in Table \ref{tab:performance_ap} REINVENT showed deteriorated performance when jointly trained with filter scores, in terms of hit ratio and top 5\% scores, implying that multi-objective optimization is more difficult than explicitly constrained optimization. Such a result was consistent for all three targets. 

\begin{table}[h]
\caption{\textbf{Performance scores of the models.} The two baseline models REINVENT and MORLD that are jointly trained to maximize filter scores are noted as REINVENT w/ filter and MORLD w/ filter. Standard deviation is given in brackets.}
\label{tab:performance_ap}
% \vspace{-0.2in}
\centering
\begin{tabular}{lll}
\toprule
          & hit ratio & top 5\% score        \\
          \midrule
MORLD     & 1.1\% (0.3\%) & -8.353 (.105) \\ 
MORLD w/ filter & 1.8\% (0.6\%) & -8.483 (.148) \\ 
REINVENT & 11.8\% (2.6\%)  & -9.649 (.186)   \\
REINVENT w/ filter & 8.8\% (2.5\%) &  -9.391 (.160)  \\
Ours: FREED(PE) & \textbf{26.3\%} (6.5\%) & \textbf{-10.426} (.314) \\
\bottomrule
\end{tabular}
\end{table}

To show that the larger library allows our model to generate more unique molecules, we provide quality scores of our model (FREED(PE)) trained with the small library and the large, unfiltered library in Table \ref{tab:quality_2_ap} and Table \ref{tab:quality_3_ap}.  

\begin{table}[h]
\caption{\textbf{Quality scores of our model (FREED(PE)) trained with the small library (number of fragments = 66).} We trained our model with all three targets and computed quality scores of the first 3,000 molecules generated during training. Standard deviation is given in brackets.}
\label{tab:quality_2_ap}
% \vspace{-0.2in}
\centering
\begin{tabular}{llllll}
\toprule
	&Glaxo	&SureChEMBL	&PAINS	&validity	&uniqueness \\
\midrule
fa7	&0.996 (0.001)	&0.808 (0.049)	&0.991 (0.002)	&1.000 (0.000)	&0.723 (0.135) \\
parp1	&0.995 (0.002)	&0.854 (0.050)	&0.991 (0.010)	&1.000 (0.000)	&0.557 (0.141) \\
5ht1b	&0.995 (0.005)	&0.823 (0.106)	&0.990 (0.007)	&1.000 (0.000)	&0.592 (0.243) \\
\bottomrule
\end{tabular}
\end{table}

\begin{table}[h]
\caption{\textbf{Quality scores of our model (FREED(PE)) trained with the large library (number of fragments = 91).} We trained our model with all three targets and computed quality scores of the first 3,000 molecules generated during training. Standard deviation is given in brackets.}
\label{tab:quality_3_ap}
% \vspace{-0.2in}
\centering
\begin{tabular}{llllll}
\toprule
	&Glaxo	&SureChEMBL	&PAINS	&validity	&uniqueness \\
	\midrule
fa7	&0.754 (0.072)	&0.558 (0.094)	&0.623 (0.201)	&1.000 (0.000)	&0.914 (0.162) \\
parp1	&0.717 (0.167) &0.577 (0.160)	&0.490 (0.169)	&1.000 (0.000)	&0.827 (0.160) \\
5ht1b	&0.690 (0.204)	&0.551 (0.200)	&0.484 (0.138)	&1.000 (0.000)	&0.801 (0.191) \\
\bottomrule
\end{tabular}
\end{table}

\paragraph{Significance analysis of the ablation study results.}  

\begin{table}[h]
\caption{\textbf{Significance analysis of the results presented in Figure \ref{fig:bar2} in the main text.} One-tail paired t-tests were performed with the null hypothesis of $m_1 \leq m_2$, where $m_1$ is the score from the compared model and $m_2$ is the score from the vanilla SAC model.}
\label{tab:significance_ap}
% \vspace{-0.2in}
\centering
\begin{tabular}{llllll}
\toprule
          & PER(PE) & PER(BU) & PER(PD) & curio(PE) & curio(BU) \\
          \midrule
fa7     & 0.00510 & 0.03410 & 0.65375 & 0.00910  & 0.30790 \\ 
parp1 & 0.00060 & 0.05000 & 0.09800 & 0.18900 & 0.39950    \\
5ht1b & 0.00660 & 0.02480 & 0.00713 & 0.13550 & 0.06150 \\
\bottomrule
\end{tabular}
\end{table}

We performed one-tail paired t-tests on the results presented in Figure \ref{fig:bar2} in the main text. We compared the models' performances in terms of hit ratio with the vanilla SAC model. The result is shown in Table \ref{tab:significance_ap}. From the $p < 0.05$ standard, we can see that FREED(PE) and FREED(BU) show significantly better results compared to the vanilla SAC model. Notably, the comparison with FREED(PE) results in a p-value of 0.0066, even below the 0.01 standard.  

\begin{figure}[h]
% \captionframework{width=.5\textwidth}
  \centering
  \includegraphics[width=.6\linewidth]{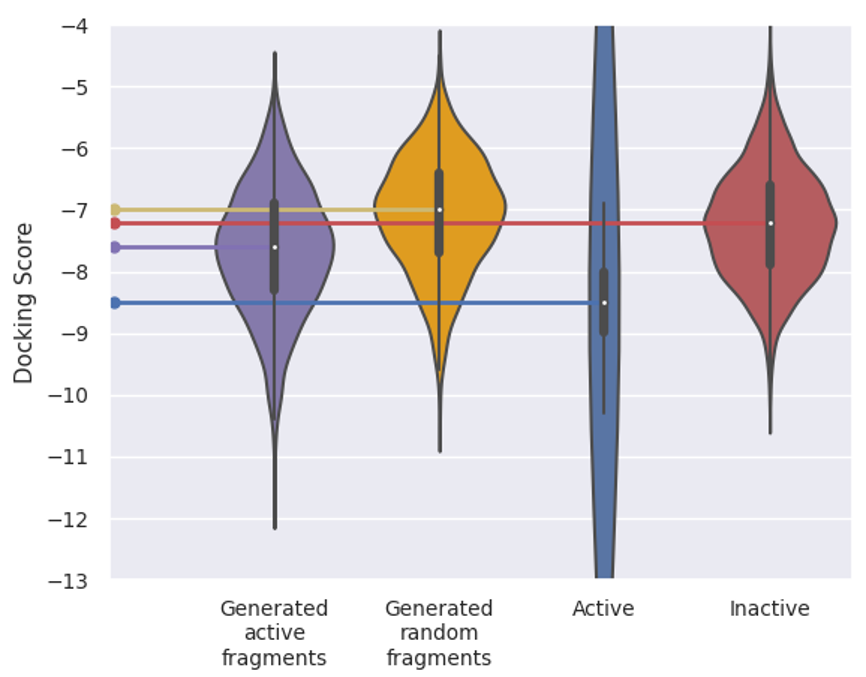}
  \caption{\textbf{Docking score distributions of the molecules generated with our method}. Molecules randomly generated from ``active fragments" and ``random fragments" are depicted as purple and yellow, respectively. Known active compounds and inactive decoys from DUD-E are depicted as blue and red, respectively.}
  \label{fig:active_ap}
% \end{wrapfigure}
\end{figure}

\paragraph{Fragments from known active compounds.}
Another advantage of our fragment-based generation method is that we can select our own fragment library. In this experiment, molecules were randomly generated until the 2,000th iteration step of our generation algorithm. Molecules were constructed from two different sets of fragments -- ``active" fragment library (see Figure \ref{fig:lib3}), a set of fragments extracted from known DUD-E active compounds of fa7 target, and ``large" fragment library, a set of fragments extracted from random ZINC molecules which were also used for experiments in Section 4.3 and Section 4.4. In Figure \ref{fig:active_ap}, active fragments-generated molecules show higher docking scores than random fragments-generated molecules. Also, while the average hit ratio was 7.24\% for random fragments-generated molecules, the average hit ratio of active fragments-generated molecules was 18.48\%, meaning that finding the potential hits becomes more than two times easier just by reconstructing molecules from the fragments of the known active compounds. 

\paragraph{Docking pose analysis of generated molecules}
In this analysis, we compare the 3D docking poses of the scaffolds and the generated leads in the Figure \ref{fig:genmol} of the main text. The 3D PyMOL \cite{PyMOL} images describe fa7 (left), parp1 (middle), and 5ht1b (right) binding with their scaffolds and the generated molecules based on those scaffolds.

\begin{figure}[h]
  \centering
  \includegraphics[width=.8\linewidth]{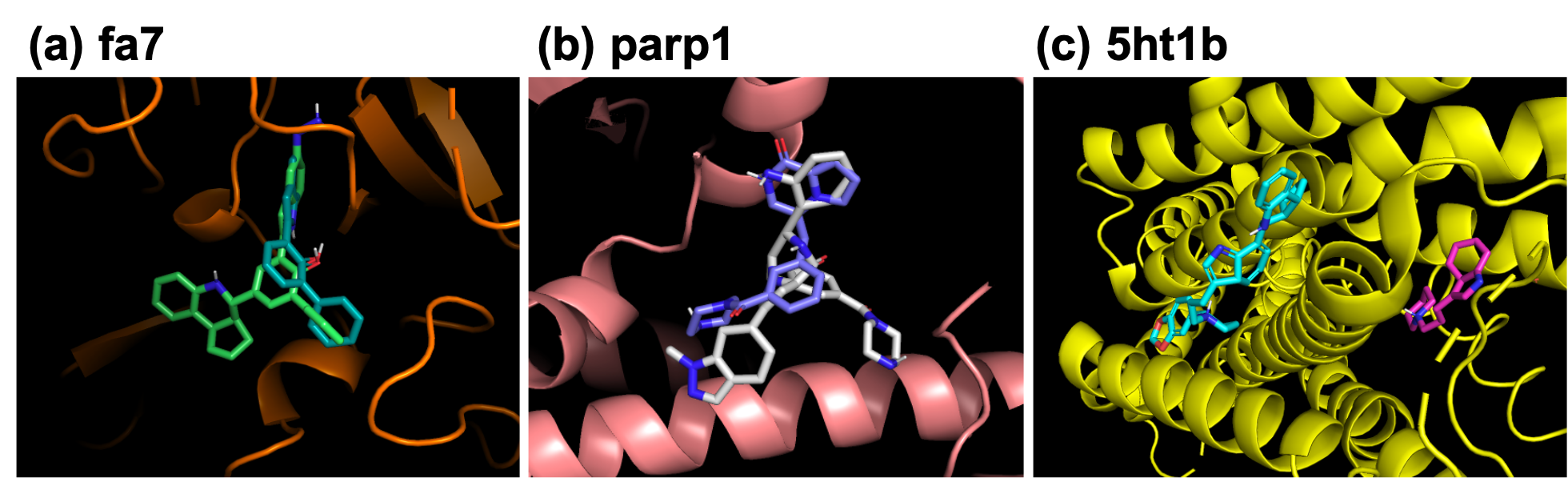}
  \caption{\textbf{3D PyMOL images of the binding poses of the scaffolds and the generated lead molecules.}}
  \label{fig:all_pose_ap}
\end{figure}

Firstly, for fa7, the binding poses of the scaffold and the generated molecules almost overlap. Such an overlap implies that the generated molecule will have a high binding affinity in high confidence. We profiled the details of the protein-ligand interactions of the scaffold and the generated molecule with a popular tool PLIP \cite{adasme2021plip}.  

\begin{figure}[h]
  \centering
  \includegraphics[width=.8\linewidth]{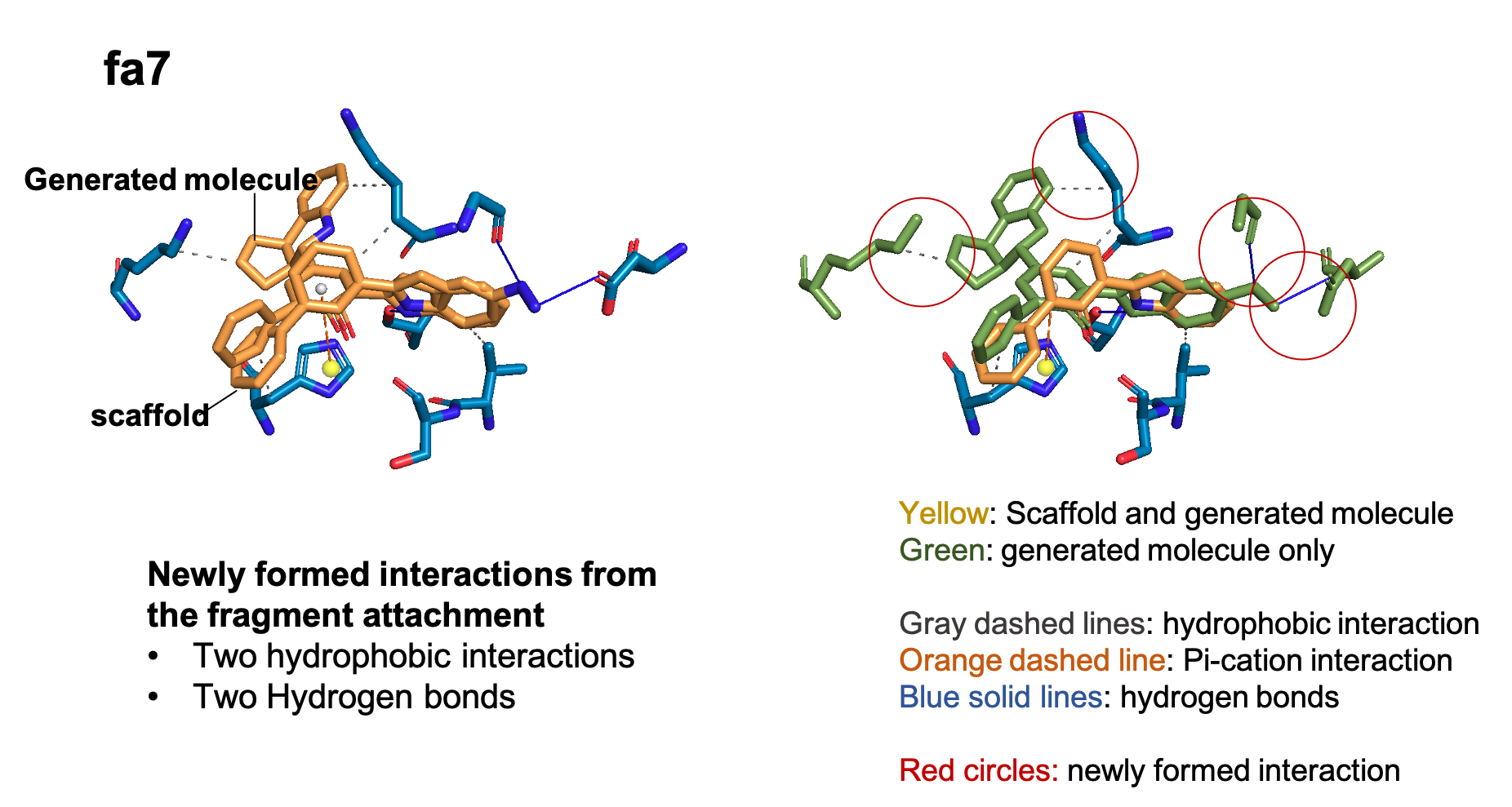}
  \caption{\small \textbf{PLIP image of the binding pose of the fa7 scaffold and the generated lead molecule.}}
  \label{fig:fa7_pose_ap}
\end{figure}

All existing interactions in the scaffold were preserved for the scaffold-based generated molecule, while several interactions were additionally formed. In detail, two hydrophobic interactions and two hydrogen bonds were formed by the augmented fragments.

For parp1, the scaffold and the generated molecule do exist in the same binding site, but the generated molecule has rotated about 180 degrees from the scaffold. Lastly, for 5ht1b, the scaffold and the generated molecule are docked in different binding sites. Since the generated molecule of 5ht1b is twice the size of the 5ht1b scaffold, we assume that the generated molecule could not fit in the original binding pocket.

Those three examples of generated molecules were randomly chosen among the high-scoring generations. Coincidentally, the three examples show us the possible orientation and binding site changes cases in scaffold-based generations. Since the changes in orientation and binding site affect the fidelity of the optimization, preventing such changes would be the future direction of our work. 

\paragraph{Scalability experiments.}
In this experiment, we tested our model's performance on the larger action space. We constructed a fragment library of 350 fragments and a fragment library of 1k fragments and trained our model on both libraries. Table shows the performance and uniqueness score.   

\begin{table}[h]
\caption{\textbf{Performance scores of the models using larger fragment libraries.} The number of fragments in the libraries is denoted in parenthesis. Standard deviation is given in brackets.}
\label{tab:scalability_ap}
% \vspace{-0.2in}
\centering
\begin{tabular}{llll}
\toprule
         & & hit ratio & top 5\% score        \\
          \midrule
\textbf{fa7} & FREED(PE) (n=91)     & 26.3\% (6.5\%) & -10.426 (.314) \\      
& FREED(PE) (n=350)     & 9.3\% (0.5\%) & -9.179 (.053) \\ 
& FREED(PE) (n=1k)     & 10.1\% (0.7\%) & -9.259 (.053) \\ 
& MORLD    & 1.1\% (0.3\%) & -8.353 (.105) \\ 
& REINVENT     & 11.8\% (2.6\%) & -9.649 (.186) \\ 
&HierVAE(AL)    & 5.5\% (0.2\%) & -9.598 (.223) \\ 
\midrule
\textbf{parp1} & FREED(PE) (n=91) & 46.4\% (8.9\%) & -13.537 (.219) \\
&FREED(PE) (n=350) & 30.5\% (1.2\%) & -11.788 (.046) \\
&FREED(PE) (n=1k) & 31.3\% (1.4\%) & -11.788 (.071) \\
&MORLD    & 5.2\% (0.9\%) & -10.551 (.143) \\ 
&REINVENT     & 29.0\% (0.3\%) & -12.565 (.298) \\ 
&HierVAE(AL)    & 25.6\% (0.9\%) & -12.306 (.414) \\ 
\midrule
\textbf{5ht1b} &FREED(PE) (n=91) & 42.1\% (12.1\%)  & -13.196 (.139)   \\
&FREED(PE) (n=350) & 57.9\% (0.4\%)  & -11.746 (.048)   \\
&FREED(PE) (n=1k) & 60.3\% (0.5\%) & -11.751 (.023) \\
&MORLD    & 11.8\% (1.1\%) & -9.715 (.073) \\ 
&REINVENT     & 45.4\% (0.5\%) & -12.094 (.301) \\ 
&HierVAE(AL)    & 13.6\% (0.4\%) & -11.779 (.353) \\ 
\bottomrule
\end{tabular}
\end{table}

Although our model achieved lower performance scores with larger fragment libraries, the scores are better than or comparable to the baseline models. 
Moreover, as expected, the models trained with the larger libraries achieved much higher uniqueness with larger fragment libraries. 
Further investigation on the appropriate size of the fragment library considering the chemical space coverage and model performance would help the practitioners, and we leave this as our future work. 

\paragraph{Distributional analysis of the generated molecules.}
In our GitHub repository \url{https://github.com/AITRICS/FREED}, we provide 2-dimensional uniform manifold approximation and projection (UMAP) analysis on the Morgan fingerprints of the generated molecules.

\subsection{Algorithm implementations}
\label{sec:A2}
This section refers to the original works of the algorithms we have employed and elaborates on our implementation.     

\paragraph{Soft Actor-Critic.}
We employ SAC as implemented in OpenAI \textit{spinningup}\footnote{MIT License, Copyright (c) 2018 OpenAI (http://openai.com)} \cite{SpinningUp2018}. Each action is predicted by corresponding policy network. 

\paragraph{Multiplicative Interaction.}
Jayakumar et al. \cite{jayakumar2019multiplicative} proposed multiplicative interaction (MI) as a powerful method to fusing information from multiple streams and showed that MI could be an alternative to basic concatenation calculation with better performances. 
We replace basic concatenation operation into MI in the case when fusing two vectors from a different scope. 
For example, MI is used in our policy network when fusing information from a node embedding vector and a graph embedding vector, in order to express the node vector's information with respect to the given graph vector.

\paragraph{Gumbel-softmax.}
In terms of implementation, sampling a single action does not require gradient flow in the original SAC. 
However, in our policy network, the actions are autoregressively defined. In detail, sampling Action 1 gives us the embeddings that will be used as an input for Action 2 policy, and likewise, sampling Action 2 gives the embeddings for Action 3 policy. Action 1, 2, and 3 together are considered a transition since the next state can only be reached after performing all three actions. As a continual gradient should flow through those autoregressive actions, we employed the Gumbel-softmax reparameterization trick \cite{jang2016categorical} to replace argmax operations.

We modify the original Gumbel-softmax formula with an additional ratio multiplied to Gumbel distribution as $\nu$:
\begin{equation}
\label{eqn:gumble_softmax}
y_i = \frac{\exp((\log(\pi_i) + \nu \cdot g_i)/ \tau)}{\sum_{j=1}^{k}\exp((\log(\pi_j)+\nu\cdot g_j)/\tau)},\\
g_i\sim \text{Gumbel}(0,1).
\end{equation}
$\nu$ is set to $10^{-3}$ and $\tau$ is set to $10^{-1}$.

\paragraph{Prioritized experience replay.}
Thrun et al. \cite{thrun1992role} and Oh et al. \cite{oh2018self} argue that exploiting important and meaningful experiences would improve the performance in difficult exploration problems.
We leverage prioritized experience replay(PER) method \cite{schaul2015prioritized} to encourage exploration in our framework.    

PER is a method that prioritizes experiences by replaying the important transitions more frequently. Probability of sampling transition $i$ is defined as 
\begin{equation}
\label{eqn:prioritized_sampling_ap}
P(i) = \frac{p_i^{\alpha}}{\sum_{k}p_k^{\alpha}}
\end{equation}
where $p_i>0$ is the priority of $i$th transition, and the exponent $\alpha$ controls how much prioritization is used.
In order to handle bias introduced from prioritized sampling, importance sampling(IS) weight $w_i$ is defined as
\begin{equation}
\label{eqn:importance_sampling}
w_i = (\frac{1}{N}\cdot\frac{1}{P(i)})^{\beta}
\end{equation}
and is multiplied to the loss defined in SAC. 

\paragraph{Curiosity-driven learning.}
Curiosity-driven exploration methods direct the agent towards exploratory directions by introducing intrinsic reward that reflects on the `surprisal' or novelty of a state. The intrinsic reward is then added to the external reward given by the environment to provide the total reward for the model update.  

Intrinsic reward of the state is defined as absolute value of predictive error of the reward predictor $r_{i}^{\text{intr}} = |\hat{y_i} - r_i|$. 
The reward predictor is separately optimized to minimize MSE loss of predicted value and actual reward value:
\begin{equation}
\label{eqn:intrinsic_loss}
\mathcal{L}_{\text{intr}}(\theta) = \sum_{i}(\hat{y_i} - r_i)^{2}
\end{equation}

\paragraph{Bayesian uncertainty.}
Kendall et al. \cite{kendall2017uncertainties} presented a Bayesian deep learning method estimating epistemic and aleatoric uncertainty. 
Monte-Carlo dropout \cite{gal2016dropout} is used to produce the mean and variance of the predicted value of the given $i$-th input. 
Predictive uncertainty for a data point $y$ can be approximated by:
\begin{equation}
\label{eqn:bayesian_uncertainty}
\text{Var}(y_i) \approx \frac{1}{T}\sum_{t=1}^{T}\hat{y}_{i,t}^{2} - (\frac{1}{T}\sum_{t=1}^{T}\hat{y}_{i,t})^{2} + \frac{1}{T}\sum_{t=1}^{T}\hat{\sigma}_{i,t}^{2}
\end{equation}
with $\{ \hat{y}_{i,t} \}_{t=1}^{T}$ as a set of $T$ sampled outputs, where $\hat{y}_{i,t}, \hat{\sigma}_{i,t}^{2} = \boldsymbol{f}^{\hat{W}_{t}}(x_i)$ for random dropout weights $\hat{W}_{t} \sim q(W)$. Estimated uncertainty in \eqref{eqn:bayesian_uncertainty} is known to capture epistemic and aleatoric uncertainty. The loss function for the Bayesian neural network is defined as:
\begin{equation}
\label{eqn:mcdropout_loss}
\mathcal{L}_{\text{BNN}}(\theta) = \frac{1}{D}\sum_{i}^\frac{1}{2\hat{\sigma}_{i}^{2}}\|y_{i} - \hat{y}_{i}\|^{2} + \frac{1}{2}\log\hat{\sigma}_{i}^{2}.
\end{equation}

\subsection{Implementation details}
\label{sec:A3}
In this section, we elaborate on the specifics of model implementation. In particular, we describe our methods of molecule digitization, fragment representation, encoder structure, and training schemes. 

\paragraph{Molecular representation.}
\begin{table}[h]
\caption{Atom features}
\label{tab:atom_features}
\centering
\begin{tabular}{ll}
\toprule
Types of atoms       & C, N, O, S, P, F, I, Cl, Br, * \\
\midrule
GetDegree()          & 0,1,2,3,4,5                    \\
GetTotalNumHs()      & 0,1,2,3,4                      \\
GetImplicitValence() & 0,1,2,3,4,5                    \\
GetIsAromatic()      & Boolean                       
\end{tabular}
\end{table}

We present features used in representing atoms in Table \ref{tab:atom_features}, where '*' is a mark for the attachment site. We used RDkit \cite{rdkit} to extract the above features from given molecules. Node features of an atom are one-hot encoded and concatenated into a vector, which is then mapped through a trainable embedding layer to produce a dense vector.

\paragraph{Fragment representation.}
In order to create representations for fragments, we utilize the Morgan circular molecular fingerprint bit vector of size 1024 and radius 2 as implemented in RDKit with default invariants that use connectivity information similar to those used for the ECFP fingerprints. Morgan fingerprints are a commonly used rule-based representation for molecules. 

\paragraph{Encoder.}
We used 3 layers of GCN \cite{kipf2016semi} with ReLU activation between each layer. Node feature vectors are first mapped to a dense vector with a linear layer. A graph with dense node vectors $H^{(0)}$ and an adjacency matrix $A$ are then mapped through a GCN layer:
\begin{equation}
\label{eqn:graph_convolutional_network}
H^{(l+1)} = \text{AGG}(\text{ReLU}(\{\Tilde{D}^{-1/2}\Tilde{A}\Tilde{D}^{-1/2}H^{(l)}W^{(l)}\}))
\end{equation}
where $\Tilde{A}=A+I$, $\Tilde{D}_{ii}=\sum_{j}\Tilde{A}_{ij}$.
Graph embedding vector is acquired by readout operation on the final (the third) node embedding vectors, i.e. $H^{(3)}$. 
We only used sum operation for aggregation (AGG) and readout, considering graph isomorphism test in molecular graphs \cite{xu2018powerful, hwang2020comprehensive}.

\paragraph{Training.}
We have set a maximum number of actions in an episode as four in \textit{de novo} generation and two in a scaffold-based generation, considering the distribution of the number of fragments in DUD-E compounds  \cite{naderi2016graph}. 
Figure 4 of \cite{naderi2016graph} shows that the most popular number of fragments in DUD-E active compounds are around 4\textasciitilde6. Accordingly, we only considered four-action episodes in this work. We did not make our algorithm learn when to stop, but we leave this as our imminent future work. 

With 64-dimensional embedding as default, we train our policy network, Q function network, graph encoder, $\alpha$ in SAC, and priority predictor using the Adam optimizer \cite{kingma2014adam},
an initial learning rate of 1e-3, weight decay of 1e-4, and a batch size of 256. Learning rates are reduced with ReduceLROnPlateau on PyTorch, with reduce factor 0.1 and patience steps 768 for policy and Q function, and 500 for priority predictor. 
In case of SAC, initial soft actor-critic $\alpha$ values are set to 1 and [min, max] value set to [0.05, 20]. The learning rate of $\alpha$ in soft actor-critic was initialized with 5e-4.
Hyper-parameters for PER follow original setting in  \cite{schaul2015prioritized}, with $\alpha$ in \eqref{eqn:prioritized_sampling} set to 0.6, and $\beta$ in \eqref{eqn:importance_sampling} set at $min(1.0, \beta_{init} + idx * (1.0 - \beta_{init}) / \beta_{frames})$, where $\beta_{init} = 0.4$ and $\beta_{frames} = 1e+5$. 
Every trainable parameter is initialized with Xavier uniform initializer \cite{glorot2010understanding}. 

\paragraph{Random exploration} In order to encourage exploration, we let the model randomly generate the experience during the first 4,000 iterations. After the random exploration, the model generates experience by the policy. The random exploration has significantly improved the model performances. 

\paragraph{Computational resources.} 
We used Intel®Xeon®Silver4210 for CPU computation including docking score calculation, and NVIDIA TitanRTX for GPU computation. 
Each random seed was run on one GPU resource, where 5 random seeds shared one CPU resource for computation.

\subsection{Experimental settings}
\label{sec:A4}
In this section, we elaborate on our experimental settings and procedures. 
 
\paragraph{Scoring function.}
We use the docking program as a scoring function to compute the final state reward. While Autodock Vina \cite{vina} is the most popular docking program for virtual screening, we used QuickVina 2 instead of AutoDock Vina to increase the speed of docking computation since QuickVina 2 is known to provide 20.39-fold acceleration compared to AutoDock Vina. Also, the binding affinity predictions from QuickVina 2 and AutoDock Vina show Pearson's correlation coefficient of 0.911 \cite{alhossary2015qvina}. Since 0.911 indicates a high correlation, we believe that using QuickVina 2 instead of AutoDock Vina would not seriously harm the fidelity.

The speed and accuracy of the docking simulation depend on its parameters -- exhaustiveness in particular. While speed and accuracy are in a trade-off relationship, we choose a high speed, low accuracy setting ($\text{exhaustiveness}=1$) to minimize the computational cost of model training. 

\begin{table}[h]
\caption{Docking configuration}
\label{tab:docking configuration}
% \vspace{-0.2in}
\centering
\begin{tabular}{ll}
\toprule
          & Configuration        \\
          \midrule
exhaustiveness     & 1  \\ 
subprocess & 10 \\ 
cpu per subprocess & 1  \\
modes & 10  \\
timeout (gen3d) & 30 sec \\
timeout (docking) & 100 sec \\
\bottomrule
\end{tabular}
\end{table}

We report the docking configuration we used for the experiments in Table \ref{tab:docking configuration}. With our computational setting, docking calculation costs around 0.9 sec per sample. 

\paragraph{Protein targets.}
Three protein targets \textbf{fa7 (FA7)}, \textbf{parp1 (PARP-1)}, \textbf{5ht1b (5-HT1B)} are chosen as design objectives to train the generative model on. 
While the model performance can greatly deviate according to its protein target, we carefully chose the targets to avoid bias in the experiments. 
The three targets have one of the highest AUROC scores when the protein-ligand binding affinities for DUD-E+ ligands are approximated with AutoDock Vina and the result was compared with ground truth, meaning that AutoDock Vina works fairly well for those three targets  \cite{cleves2020auroc}. We assume that QuickVina 2, a derivative of AutoDock Vina, would similarly work well for the three targets. 
Also, three targets have different protein family memberships - fa7 (Coagulation factor VII) in protease family, parp1 (Poly [ADP-ribose] polymerase-1) in polymerase family, and 5ht1b (5-hydroxytryptamine receptor 1B) in G protein-coupled receptor family.

\paragraph{\textit{De novo} and scaffold-based scenario.} 
\textit{De novo} drug design is to generate molecules with high therapeutic potential from scratch. Scaffold-based drug design is to add or modify substructures from the given scaffold. In our experiments, we prepare a benzene ring as an initial molecule from \textit{de novo} scenario since aromatic rings are common substructures in druglike molecules.   

For scaffold-based scenarios, we detect scaffold for each target by the procedure described below and use those scaffolds as initial molecules. Note that our model requires the user to set the number of fragments to add to the initial molecule. We allow four augmentation steps for \textit{de novo} designs and two augmentation steps for scaffold-based designs. 

\textbf{Scaffold detection.} We detect scaffold of DUD-E \cite{mysinger2012directory} (fa7, parp1) or ChEMBL \cite{gaulton2012chembl} (5ht1b) active compounds using MurckoScaffold function implemented by RDKit \cite{rdkit}. 
Then, we sort the scaffolds by frequency and take the most frequently observed scaffold as an initial molecule. We observed several active compounds which include chosen scaffold as their substructures, and carefully chose the attachment sites that connect the scaffold to surrounding atoms.

\begin{figure}[h]
\hspace*{-0.1in}
  \centering
  \includegraphics[width=1.0\linewidth]{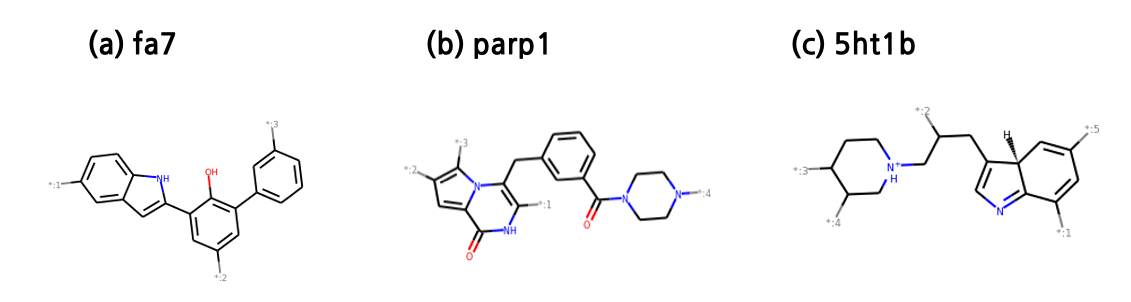}
  \caption{\small Detected scaffolds of fa7, parp1, and 5ht1b}
  \label{fig:scaff}
\end{figure}

\paragraph{Pharmacochemical filters.}   
We utilize widely-used pharmacochemical filters to assess the generated molecule's quality. We calculate the ratio of molecules accepted by each filter to total generated molecules and report the result in Section 4.2 Table 1. The filters are also used to exclude inappropriate fragments from the fragment library.  

\textbf{1) PAINS}: Pan-assay interference compounds (PAINS) \cite{baell2010new} are chemical compounds that tend to bind nonspecifically with numerous biological targets rather than discriminately affecting one desired target, often giving false positive results in high-throughput screening \cite{dahlin2015pains, baell2014pains}. The PAINS filter contains 481 structural alerts.  

\textbf{2) SureChEMBL Non MedChem-Friendly SMARTS}: SureChEMBL Non MedChem-Friendly SMARTS \cite{sushko2012toxalerts} is a set of structural alerts or toxicophores, which are substructures that are highly correlated with properties undesirable for drugs typically associated with human or environmental toxicity. When such structural alerts are used to filter medicinally unfriendly compounds in early-stage drug discovery, a significant reduction in compound failure rates in the clinic has been observed. The SureChEMBL Non MedChem-Friendly SMARTS contains 166 structural alerts \cite{papadatos2016surechembl}. 

\textbf{3) Glaxo Hard Filters}: Glaxo Hard Filters are a set of substructure filters that rejects compounds containing inappropriate functional groups, such as reactive functional groups, unsuitable leads (i.e., compounds which would not be initially followed up), and unsuitable natural products (i.e., derivatives of natural product compounds known to interfere with common assay procedures). The Glaxo Hard Filters contain 51 structural alerts \cite{hann2016glaxo}.

\begin{figure}[]
\hspace*{-0.1in}
  \centering
  \includegraphics[width=1.0\linewidth]{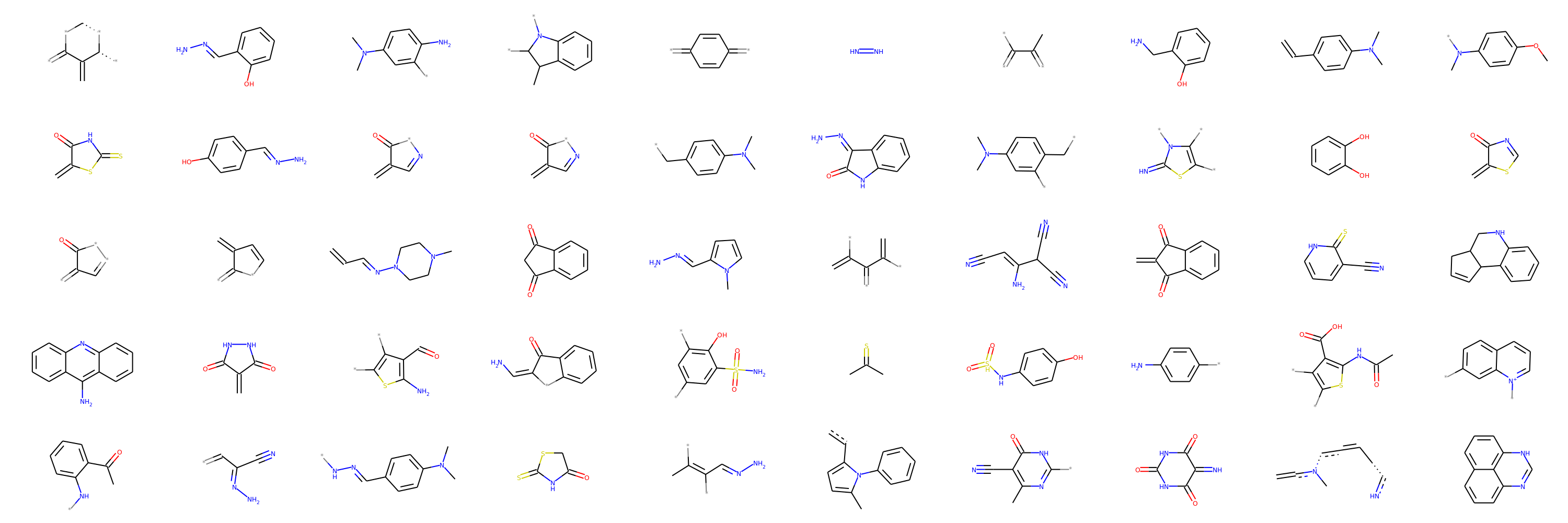}
  \caption{\small 50 examples of structural alerts in PAINS filter}
  \label{fig:pains}
\end{figure}

\begin{figure}[]
\hspace*{-0.1in}
  \centering
  \includegraphics[width=1.0\linewidth]{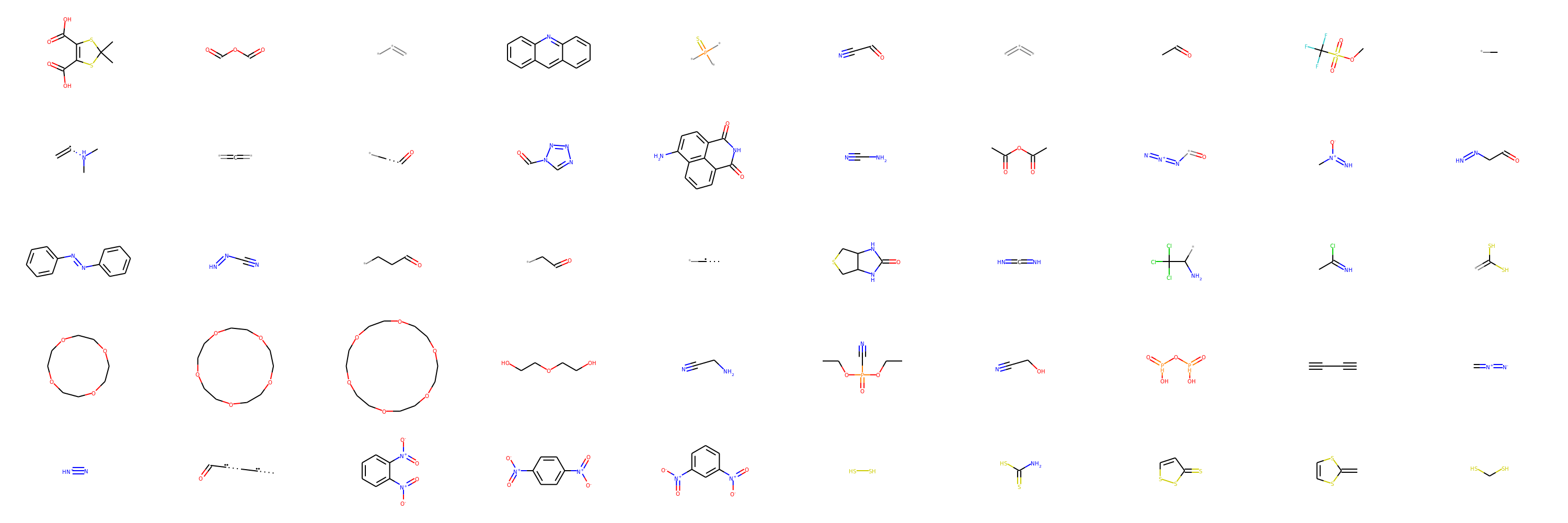}
  \caption{\small 50 examples of structural alerts in SureChEMBL Non MedChem-Friendly SMARTS}
  \label{fig:sure}
\end{figure}

\begin{figure}[]
\hspace*{-0.1in}
  \centering
  \includegraphics[width=1.0\linewidth]{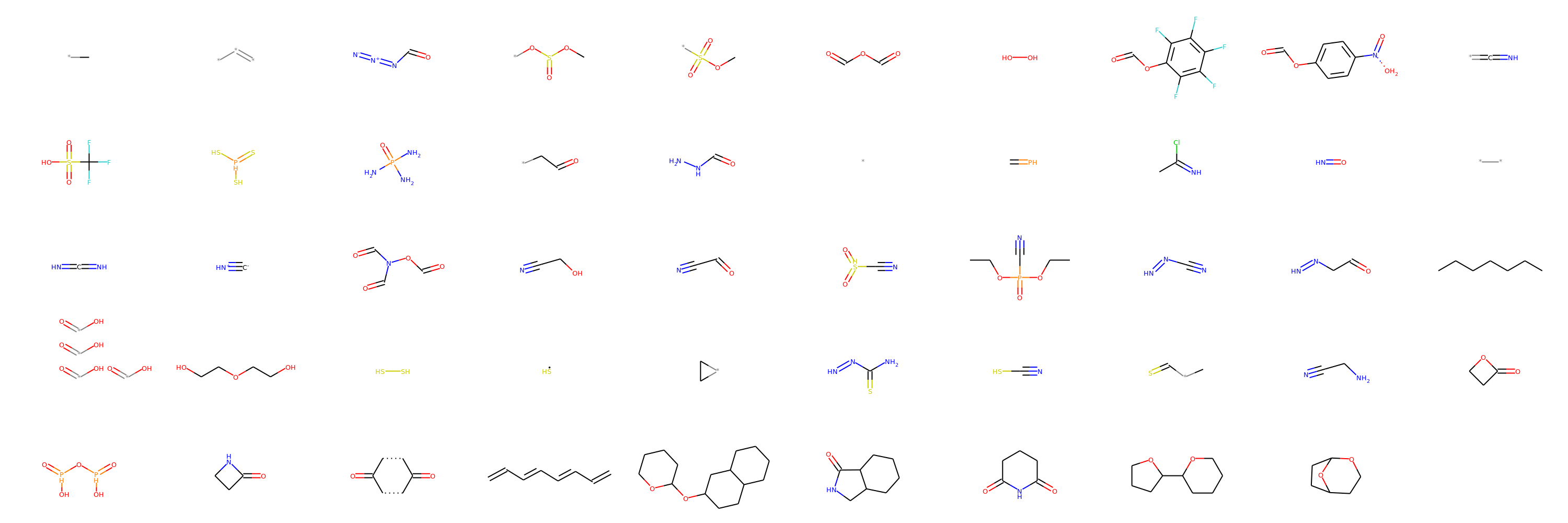}
  \caption{\small 49 examples of structural alerts in Glaxo Hard Filter}
  \label{fig:glaxo}
\end{figure}

\paragraph{Fragment library.} 

For \textbf{random fragments}, we fragmented 250k druglike molecules in the ZINC database \cite{irwin2005zinc} and filtered the fragments according to a number of atoms, radius, and frequency. 
We only took the fragments that contained fewer than 12 atoms, and we excluded fragments that appear only once or twice in the ZINC database. 

\begin{figure}[]
\hspace*{-0.1in}
  \centering
  \includegraphics[width=1.0\linewidth]{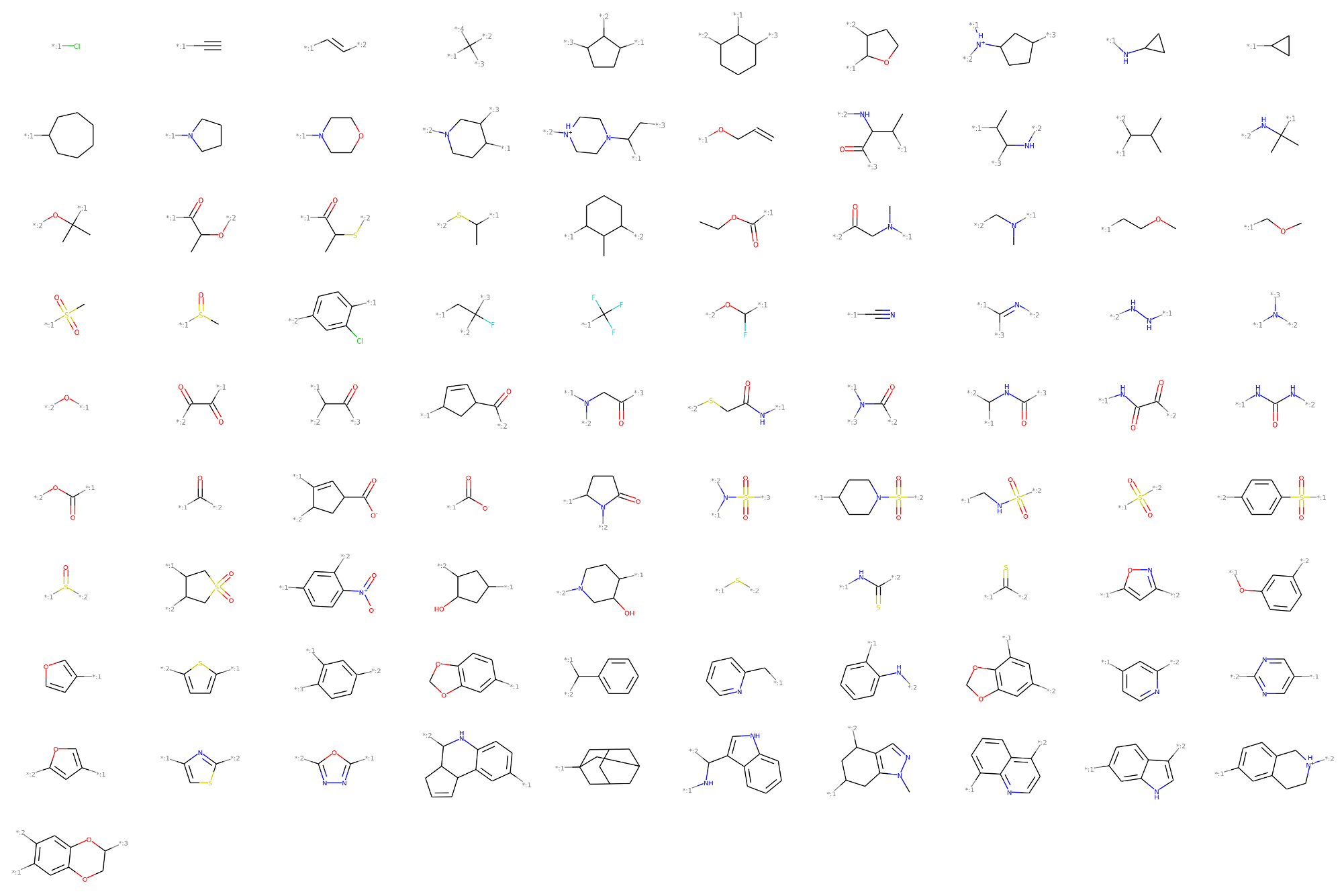}
  \caption{\small 91 fragments in \textbf{large library}.}
  \label{fig:lib1}
\end{figure}

\begin{figure}[]
\hspace*{-0.1in}
  \centering
  \includegraphics[width=1.0\linewidth]{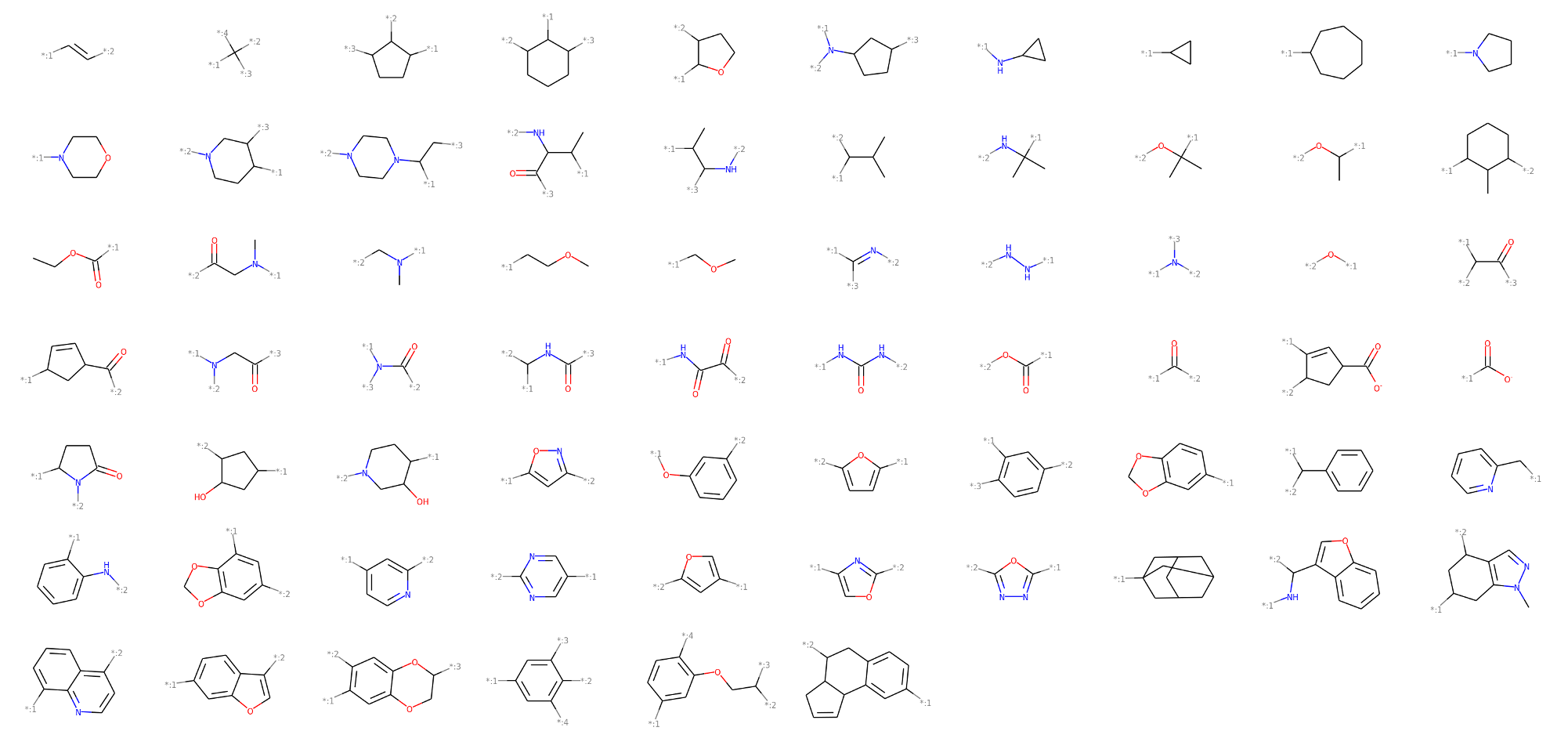}
  \caption{\small 66 fragments in \textbf{small library}.}
  \label{fig:lib2}
\end{figure}

\begin{figure}[]
\hspace*{-0.1in}
  \centering
  \includegraphics[width=1.0\linewidth]{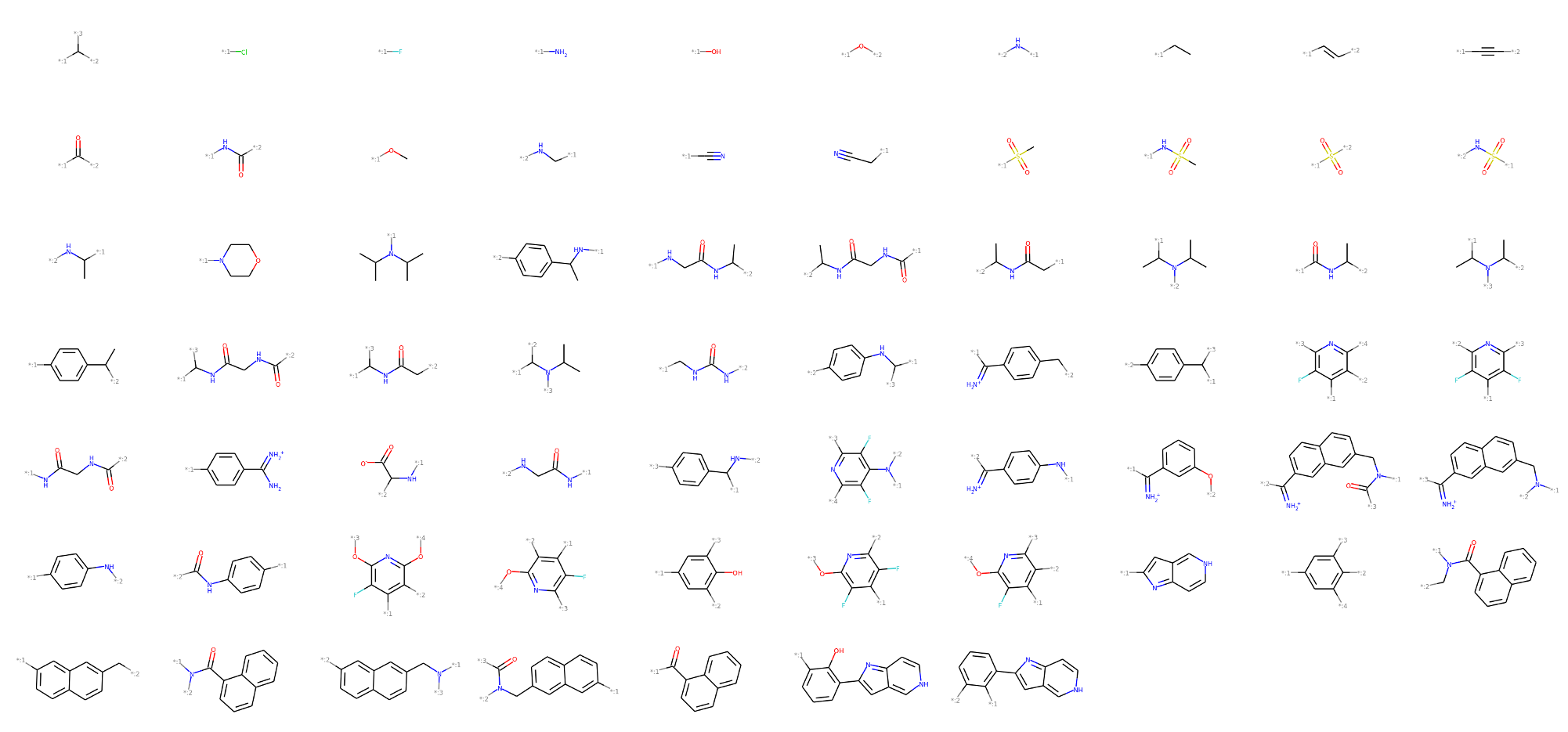}
  \caption{\small 67 fragments in \textbf{active library}.}
  \label{fig:lib3}
\end{figure}

In the fragment filtering procedure, fragments that might evoke RDKit parse errors were excluded. Also, if two fragments with the same graph structure have different attachment sites, we remove the one that has fewer attachment sites.
For experiments in Section 4.3 and Section 4.4, we use \textbf{large fragment} library, which does not exclude filter-rejected fragments. We choose 91 fragments that appear most frequently in ZINC druglike molecules. 
For experiments in Section 4.2, we use \textbf{small library} where we excluded fragments rejected by PAINS, SureChEMBL, and Glaxo filters from \textbf{large library}. The \textbf{small library} consists of 66 fragments. The structures of the fragments are shown in Figure \ref{fig:lib1} and Figure \ref{fig:lib2}. 
We provide Jupyter notebook for fragment library generation in our GitHub repository \url{https://github.com/AITRICS/FREED}. 

\subsection{Baseline frameworks} 
\label{sec:A5}
We used MORLD and REINVENT as our baselines since those two models are utilized to optimize the docking score in previous works. While MORLD is an atom-based generative model, REINVENT can be considered as a representative SMILES-based generative model. We also compared our model with HierVAE. HierVAE is a strong variational autoencoder (VAE)-based method that generates molecular graphs using structural motifs as building blocks. 

\paragraph{MORLD.} MORLD \cite{jeon2020morld} is an atom-based molecular generative model guided by the MolDQN algorithm. We set a benzene ring as an initial molecule for MORLD training. We use QuickVina 2 docking program as a scoring function to train MORLD as in the original work. We jointly trained QED and SA scores as in the original work.

\paragraph{REINVENT.} REINVENT \cite{olivecrona2017reinvent} is a smiles-based molecular generative model guided by an improved REINFORCE algorithm. We used the RNN model pretrained on a 1.5 million ZINC druglike molecule dataset. The pretrained model was provided by the original work \footnote{\url{https://github.com/MarcusOlivecrona/REINVENT}}. While the Cieplinksi et al. \cite{cieplinksi2020atleast} and Bender et al. \cite{bender2021gpcr} utilized the SMINA docking program as its scoring function, we use QuickVina 2 docking program as a scoring function to reduce the computational cost.  

\paragraph{HierVAE.}
We used the model checkpoint uploaded on the official GitHub repository, which was pretrained with ChEMBL molecules \footnote{\url{https://github.com/wengong-jin/hgraph2graph}}. We also used the HierVAE vocab (fragment) library from the official GitHub. 89.6\% fragments of the vocab are Glaxo-accepted molecules, 81.8\% are SureChEMBL-accepted molecules, and 99.9\% are PAINS-accepted molecules.  

We then fine-tuned the model with respect to three protein targets (fa7, parp1, 5ht1b) using the corresponding DUD-E active molecules as the training set. After removing the molecules that evoke RDKit parse errors, we obtained as the training sets 99 samples for fa7, 433 samples for parp1, and 1129 samples for 5ht1b. We trained the models for the proteins until the loss converges. Since the 5ht1b training set has a larger size, the 5ht1b model converged more slowly than the other two, and we trained until 700 epochs for fa7 and parp1, and 800 epochs for 5ht1b. We generated 3,000 molecules with each model, and we computed the hit ratio and top 5\% score.    

For active learning scheme (HierVAE(AL)), we first fine-tuned the ChEMBL pretrained HierVAE with the active molecule set from DUD-E. Then, we generated 3,100 molecules from the fine-tuned HierVAE, computed the molecules' docking scores, and gathered the set of molecules whose docking scores are better than the active threshold (=the median of the docking scores of the active molecules), and fine-tuned the model with these molecules. We repeated the collecting of the molecules and fine-tuning twice. In this scheme, we perform \textasciitilde6,000 docking computations, which is roughly twice of the other methods.
The number of selected molecules in the first round of training were 1,138 for 5ht1b, 1,732 for parp1, and 291 for fa7. For second round, 585 molecules for 5ht1b, 1,790 for parp1, and 346 for fa7 were used for training. The  model performance was improved in terms of both hit ratio and top 5\% score, compared to the one-time training scheme.  

% \bibliography{refer_appendix}

\end{document}